\documentclass[12pt,fleqn]{article}

\usepackage{booktabs}
\usepackage{doi}
\usepackage{graphicx}
\usepackage{amsmath}
\usepackage{amssymb,amsmath}
\usepackage{mathrsfs}
\usepackage{color}
\usepackage[margin=1 in]{geometry}

\usepackage{enumitem}
\usepackage{hyperref}

\numberwithin{equation}{section}

\newcommand{\pd}{\partial}
\newcommand{\cphi}{\varphi}

\newcommand{\pderA}[2]{\ensuremath{\frac{\partial #1}{\partial #2}}}

\usepackage{bm}

\begin{document}
	
\title{New classes of quadratically integrable systems in magnetic fields: the generalized cylindrical and spherical cases}
\author{O. Kub\r{u}$^1$, A. Marchesiello$^2$ and L. \v{S}nobl$^1$}
\maketitle

\begin{center}
{$^1$Czech Technical University in Prague, Faculty of Nuclear Sciences and Physical
	Engineering, Department of Physics, B\v{r}ehov\'{a} 7, 115 19 Prague 1, Czech Republic}\\
{$^2$Czech Technical University in Prague, Faculty of Information Technology,
	Department of Applied Mathematics, Th\'{a}kurova 9, 160 00 Prague 6, Czech Republic} \\[1mm]

E-mail: Ondrej.Kubu@fjfi.cvut.cz, marchant@fit.cvut.cz, Libor.Snobl@fjfi.cvut.cz 
\end{center}

\vspace{10pt}

\begin{abstract}
We study integrable and superintegrable systems with magnetic field possessing quadratic integrals of motion on the three-dimensional Euclidean space. In contrast with the case without vector potential, the corresponding integrals may no longer be connected to separation of variables in the Hamilton--Jacobi equation and can have  more general leading order terms.
We focus on two cases extending the physically relevant cylindrical-- and spherical--type integrals. We find three new integrable systems in the generalized cylindrical case but none in the spherical one. We conjecture that this is related to the presence, respectively absence, of maximal abelian Lie subalgebra of the three-dimensional Euclidean algebra generated by first order integrals in the limit of vanishing magnetic field. By investigating superintegrability, we find only one (minimally) superintegrable system among the integrable ones. It does not separate in any orthogonal coordinate system. This system provides a mathematical model of a helical undulator placed in an infinite solenoid.
	\end{abstract}

\vspace{2pc}
\maketitle


\section{Introduction}

We study conditions for the integrability and superintegrability of classical Hamiltonian systems in the three-dimensional (3D) Euclidean space in the presence of a magnetic field. Let us recall that a Hamiltonian system  which does not explicitly depend on time is said to be integrable if it admits a pair of constants of motion that are in involution, i.e., a pair of integrals $X_1$, $X_2$ that Poisson commute with the Hamiltonian and with each other. For superintegrability we require additional independent integral(s) (which cannot be in involution with all the others). In the following we focus on the conditions for the existence of quadratic integrals, i.e., we restrict ourselves to the case in which both integrals are (at most) second order polynomials in the momenta. 

In the absence of vector potentials, the problem on the quadratical integrability of such three dimensional Hamiltonian systems in the Euclidean space, also called systems with only scalar potentials or natural systems, has been completely solved, and a classification has been achieved \cite{Makarov1967}. Superintegrable systems with quadratic integrals have also been extensively studied and the list of superintegrable systems with two pairs of integrals in involution is known \cite{Evans}. Furthermore, integrability with quadratic integrals is equivalent to separability in a suitable orthogonal coordinate system and superintegrability (with two pairs of integrals in involution) implies separability in more than one orthogonal coordinate system. Such systems are thus also called multi--separable.
 
Here we consider the problem of integrability for the 3D Hamiltonian in a nonvanishing magnetic field. There is a variety of forms that may be admissible for the leading order terms of a pair of quadratic integrals in involution \cite{Marchesiello2022}. However, though it is known that some of such generalized forms of the integrals indeed exist in a magnetic field \cite{Marchesiello2017,Marchesiello2022}, it is not clear which of the possibilities allowed by the algebraic structure as classified in~\cite{Marchesiello2022} can yield new integrable systems with such generalized--type integrals and which, on the contrary, reduce only to the standard forms known for scalar potentials. Moreover, such classes of integrals are generically not related to separation of variables. Indeed, with magnetic fields, necessary but not sufficient condition for separability is the existence of at least one first order integral \cite{Shapovalov1972,Benenti_2001}.

From the examples known so far in two dimensions for scalar potentials we observed some disparity in the allowed forms of the leading order terms of the integrals, depending on whether the separating coordinates are subgroup type or not\footnote{The subgroup type coordinates in 3D are defined by the property that some subgroup chain
\begin{equation}
E_3 \supset \tilde{G} \supset G_M
\end{equation}
exists where $E_3$ is the Euclidean group and $G_M$ is its Abelian subgroup such that the leading order terms of their corresponding quadratic integrals of motion are second--order Casimir operators of subgroups \cite{KMW76,MPW81}.}. This concerns superintegrable systems with one quadratic and one third or higher order integral \cite{Marchesiello2015a}. 2D subgroup type coordinates are Cartesian coordinates and polar coordinates. They are related to the maximal (abelian) Lie subalgebras of the two dimensional Euclidean algebra, namely $\mathrm{span}\{p_1,p_2\}$ and $\mathrm{span}\{\ell_3\}$, respectively, where $p_j$ denotes the linear momenta and $\ell_3$ is the third component of the angular momentum.

In the presence of magnetic fields, the connection with separability is lost already when second order integrals are considered, and we wonder if the difference in behavior mentioned above already appears.

As the existence of generalized Cartesian integrals is known to be possible, we investigate here the possibility of the existence of generalized--type integrals corresponding to the other two 3D subgroup type coordinates, namely cylindrical and spherical. We find that quadratically integrable systems exist with generalized cylindrical--type integrals, but none with generalized spherical ones. Furthermore, all the systems known so far with generalized--type integrals, including the new ones presented here, would separate in Cartesian or cylindrical coordinates in the limit of vanishing magnetic field \cite{Marchesiello2017}, not in other coordinate systems, even in the case the integrals do not generalize the standard Cartesian nor cylindrical one \cite{Marchesiello2022}. The reason for this dissimilarity is to our knowledge not yet understood and could be the key to achieve a full classification of quadratically integrable systems in magnetic fields.

We also search for quadratically superintegrable systems among the found integrable ones. We find only one such system. It is also the first known non--separable superintegrable system with magnetic field on 3D Euclidean space.

The structure of the paper is as follows: In Section \ref{sec: problem} we review Hamiltonian systems with magnetic fields and the form of the corresponding quadratic first integrals. In Section \ref{sec: ext cyl} we search for integrable systems with generalized cylindrical integrals, followed by Section \ref{nonsep Cart} analyzing the new superintegrable system found among them. We analyze its Poisson algebra, reduce it to a 2D system and prove that it does not separate in any coordinate system. In Section \ref{sec:GenSpher} we exclude the existence of integrable systems of the generalized spherical type. We conclude (Section \ref{sec:con}) with some interpretation of the obtained results.

\section{The systems and their integrals of motion}\label{sec: problem}
We consider the system given by
\begin{equation}\label{HamMagn}
	H=\frac{1}{2}\left(\vec{p}+\vec{A}(\vec{x})\right)^2+W(\vec{x}),
\end{equation}
where $W(\vec{x})$ is called the electrostatic potential and, due to the magnetic field, we have also the vector potential $\vec A(\vec{x})$.

Time--independent gauge transformations
\begin{equation}\label{cl gauge}
	\vec A'(\vec{x})= \vec{A} (\vec{x})+\mathrm{d} \chi(\vec{x}), \quad W'(\vec{x})=W(\vec{x})
\end{equation}
leave the magnetic field 
\begin{equation}
	\vec{B}(\vec{x})=\nabla \times \vec{A}(\vec{x})
\end{equation} 
and electrostatic potential $W(\vec{x})$ of the system invariant. 

The scalar potential  (i.e., the momentum-free terms in the Hamiltonian~\eqref{HamMagn}) 
\begin{equation}
V(\vec x)= W(\vec x)+\frac12\left(A_1(\vec x)^2+A_2(\vec x)^2+A_3(\vec x)^2\right)
\end{equation}
is affected by the transformation \eqref{cl gauge}. First and zero order terms in momenta in the integrals can change under~\eqref{cl gauge} as well. Therefore, we find it convenient to introduce the covariant expression for the momenta
\begin{equation}
	p_j^A=p_j+A_j(\vec{x})
\end{equation} 
and consider a general second order integral expressed in the form
\begin{equation}\label{IntCart}
	X=\sum_{j=1}^3 h^j(\vec{x})p_j^Ap_j^A + \sum_{j,k,l=1}^3 \frac{1}{2}|\epsilon_{jkl}|n^j(\vec{x})p_k^Ap_l^A + \sum_{j=1}^3 s^j(\vec{x})p_j^A + m(\vec{x}),
\end{equation} 
where $h^j(\vec{x})$, $n^j(\vec{x})$, $s^j(\vec{x})$ ($j=1,2,3$) and $m(\vec{x})$ are real--valued functions.

To be an integral of motion, $X$ has to be in involution with the Hamiltonian, namely
\begin{equation}\label{Int_cond0}
\{ X, H\}=0,
\end{equation}
where $\{\;,\;\}$ denotes the Poisson bracket $\{A,B\}=\sum_{i=1}^{3} \frac{\partial A}{\partial x_j} \frac{\partial B}{\partial p_j}-\frac{\partial A}{\partial p_j} \frac{\partial B}{\partial x_j}$.
The condition~\eqref{Int_cond0} is a polynomial of third order in the momenta. By collecting the monomial terms of different order and equating their coefficients to zero, we obtain the determining equations for the integral $X$ \cite{Marchesiello2015} that, for sake of completeness, we write in the following. From the third order terms we have
\begin{align}\label{third QM cart}
	&\pd_x h^x =0, \quad \pd_y h^x =-\pd_x n^x, \quad \pd_z h^x =-\pd_x n^y, \nonumber\\
	&\pd_x h^y =-\pd_y n^z, \quad \pd_y h^y =0, \quad \pd_z h^y =-\pd_y n^x, \\
	&\pd_x h^z =-\pd_z n^z, \quad \pd_y h^z =-\pd_z n^x, \quad \pd_z h^z =0, \nonumber\\
	&\nabla\cdot \vec{n}=0.	\nonumber
\end{align}
The second order equations are
\begin{align}\label{second QM cart}
	&\pd_x s^x=n^y B^y-n^z B^z, \nonumber\\
	&\pd_y s^y=n^z B^z-n^x B^x, \nonumber\\
	&\pd_z s^z=n^x B^x-n^y B^y, \nonumber\\
	&\pd_y s^x+\pd_x s^y=n^x B^y-n^y B^y+2(h^x-h^y)B^z, \\
	&\pd_z s^x+\pd_x s^z=n^z B^x-n^x B^z+2(h^z-h^x)B^y, \nonumber\\
	&\pd_y s^z+\pd_z s^y=n^y B^z-n^z B^y+2(h^y-h^z)B^x\nonumber
\end{align}
and imply
\begin{equation}
	\nabla \cdot \vec{s}=0.
\end{equation}
From the first order terms we obtain
\begin{align}\label{first QM cart}
	&\pd_x m=2h^x\pd_x W+n^z\pd_y W+n^y\pd_z W+ s^z B^y-s^y B^z, \nonumber\\
	&\pd_y m=2h^y\pd_y W+n^z\pd_x W+n^x\pd_z W+ s^x B^z-s^z B^x, \\
	&\pd_z m=2h^z\pd_z W+n^y\pd_x W+n^x\pd_y W+ s^y B^x-s^x B^y,\nonumber
\end{align}
and finally, the zeroth order equation reads
\begin{equation}\label{zeroth QM cart}
	\begin{split}
		\vec{s}\cdot \nabla W=0.
	\end{split}
\end{equation}
The solution of the third order equations is known \cite{Miller2013} and it implies
\begin{equation}\label{int cart}
	X=\sum_{1 \leq a \leq b \leq 6} \alpha_{ab}Y_a^A Y_b^A + \sum_{j=1}^3 s^j(\vec{x})p_j^A + m(\vec{x}),
\end{equation} 
where 
\begin{equation}
	Y^A=(p_1^A,p_2^A,p_3^A,\ell_1^A,\ell_2^A,\ell_3^A), \qquad \ell_i^A=\sum_{1 \leq j,k \leq 3} \epsilon_{ijk}x_j p_k^A.
\end{equation} 
Here we address the problem of existence of a pair of integrals mutually in involution of generalized cylindrical and spherical type. From the classification of pairs of commuting quadratic elements in the enveloping algebra of the Euclidean algebra \cite{Marchesiello2022} we consider the cases (f) and (a), namely integrals of the form
\begin{equation}\label{int gen ext cyl}
	X_1=(\ell^A_3)^2-a \ell^A_3 p^A_3+b (p_1^A)^2 +(c p^A_1 +d p^A_2) p^A_3+\ldots,\qquad X_2=(p^A_3)^2+\ldots
\end{equation}
and
\begin{equation}\label{int gen spher ext}
X_1=(\ell_3^{A})^2+\dots,\qquad X_2=(\ell_1^{A})^2+ (\ell_2^{A})^2+ (\ell_3^{A})^2+ a \ell_3^{A} p_3^{A}+b(p_3^A)^2 +\dots,
\end{equation}
where $a,b, c, d$ are real parameters. For $b=0$ these generalize cylindrical and spherical type integrals, for $b\neq 0$ they can be interpreted as generalizations of non--subgroup type classes, namely elliptic cylindrical and oblate/prolate spheroidal, respectively. In this work we focus on the physically more relevant cases of generalized cylindrical and spherical type integrals (which is also computationally more feasible).

Though integrals \eqref{int gen ext cyl} and \eqref{int gen spher ext} are not in general related to separation of variables in the cylindrical and spherical coordinates, respectively, we find it convenient to work in such coordinate systems in order to solve their determining equations. Therefore, let us introduce the cylindrical
\begin{equation}\label{cyl coords}
	x=r \cos\cphi,\quad y=r \sin\cphi,\quad z=Z
\end{equation}
and spherical 
\begin{equation}\label{sph coords}
	x=R\sin\theta\cos\phi,\quad y=R\sin\theta\sin\phi,\quad z=R\cos \theta
\end{equation}
coordinates. By using
\begin{equation}
	\lambda = p_x \mathrm{d}x + p_y \mathrm{d}y + p_z \mathrm{d}z = p_r \mathrm{d}r + p_\cphi \mathrm{d}\cphi + p_Z \mathrm{d}Z = p_R \mathrm{d}R + p_\theta \mathrm{d}\theta + p_\phi \mathrm{d}\phi,
\end{equation} 
we obtain the following transformations of the components of the linear momentum as the components of a 1--form: 
\begin{align}
	p_x &= \cos\cphi p_r-\frac{\sin\phi}{r}p_\cphi, \quad p_y = \sin\cphi p_r+\frac{\cos\cphi}{r}p_\cphi, \quad p_z = p_Z; \\
\nonumber	p_x &= \sin\theta\cos\phi p_R+\frac{\cos\theta \cos\phi}{R}p_\theta-\frac{\sin \phi}{R\sin\theta}p_\phi,\\
	p_y &= \sin\theta\sin\phi p_R+\frac{\cos\theta \sin\phi}{R}p_\theta+\frac{\cos \phi}{R\sin\theta}p_\phi,\\
\nonumber	p_z &= \cos\theta p_R-\frac{\sin \theta}{R}p_\theta,
\end{align}
respectively. The components of the vector potential transform in the same way as the components of the momentum.

The components of the magnetic field 2--form $B=\mathrm{d}A$ are
\begin{align}
	\begin{aligned}
		B &= B^x (\vec{x})\,\mathrm{d}y \wedge \mathrm{d}z + B^y (\vec{x})\, \mathrm{d}z \wedge \mathrm{d}x + B^z (\vec{x})\, \mathrm{d}x \wedge \mathrm{d}y \\
		&= B^r (r, \cphi, Z)\, \mathrm{d}\cphi \wedge \mathrm{d}Z + B^\cphi (r, \cphi, Z)\, \mathrm{d}Z \wedge \mathrm{d}r + B^Z(r, \cphi, Z)\, \mathrm{d}r \wedge \mathrm{d}\cphi \\
		&= B^R (R, \theta, \phi)\, \mathrm{d}\theta \wedge \mathrm{d}\phi + B^\theta (R, \theta, \phi)\, \mathrm{d}\phi \wedge \mathrm{d}R + B^\phi(R, \theta, \phi)\, \mathrm{d}R \wedge \mathrm{d}\theta.
	\end{aligned}
\end{align}
This leads to the following transformations
\begin{align} \label{transformB}
	B^x (\vec{x}) &= \frac{\cos\cphi}{r}B^r (r, \cphi, Z) - \sin\cphi B^\cphi (r, \cphi, Z), \nonumber\\
	B^y (\vec{x}) &= \frac{\sin\cphi}{r}B^r (r, \cphi, Z) + \cos\cphi B^\cphi (r, \cphi, Z), \\
	B^z (\vec{x}) &= \frac{1}{r}B^Z (r, \cphi, Z); \nonumber \\
	B^x (\vec{x}) &= \frac{\cos\phi}{R^2}B^R (R, \theta, \phi)+\frac{\cos\theta \cos\phi}{R\sin\theta}{B^\theta (R, \theta, \phi)} - \frac{\sin\phi}{R} B^\phi (R, \theta, \phi), \nonumber\\
	B^y (\vec{x}) &= \frac{\sin\phi}{R^2}B^R (R, \theta, \phi)+\frac{\cos\theta \sin\phi}{R\sin\theta}{B^\theta (R, \theta, \phi)} + \frac{\cos\phi}{R} B^\phi (R, \theta, \phi), \label{transformBsph} \\
	B^z (\vec{x}) &=\frac{\cos\theta}{R^2\sin\theta}	{B^R (R, \theta, \phi)}- \frac{1}{R}B^\theta (R, \theta, \phi), \nonumber
\end{align}
so that the components of the magnetic field are computed in the same way as in the Cartesian coordinates, namely
\begin{align}\label{B=dA}
	 B^r &=\pderA{A_Z}{\cphi}-\pderA{A_\cphi}{Z}, \quad B^\cphi=\pderA{A_r}{Z}-\pderA{A_Z}{r}, \quad B^Z=\pderA{A_\cphi}{r}-\pderA{A_r}{\cphi}, \\
	 B^R &=\pderA{A_\phi}{\theta}-\pderA{A_\theta}{\phi}, \quad B^\theta=\pderA{A_R}{\phi}-\pderA{A_\phi}{R}, \quad B^\phi=\pderA{A_\theta}{T}-\pderA{A_R}{\theta}.
\end{align}
In the cylindrical coordinates, the Hamiltonian~\eqref{HamMagn} reads as follows
\begin{equation}\label{HamMagnCyl}
	H=\frac{1}{2}\left(\left(p_r^A\right)^2+\frac{\left(p_\cphi^A\right)^2}{r^2}+\left(p_Z^A\right)^2\right)+W(r,\cphi,Z),
\end{equation} 
where
\begin{equation}
	p_r^A = p_r + A_r (r,\cphi,Z), \quad p_\cphi^A = p_\cphi + A_\cphi (r,\cphi,Z), \quad p_Z^A = p_Z + A_Z (r,\cphi,Z).
\end{equation} 
Similarly, in the spherical coordinates the Hamiltonian reads 
\begin{equation}\label{HamMagnSpher}
H=\frac{1}{2}\left(\left(p_R^A\right)^2+\left(\frac{p_\theta^A}{R}\right)^2+\left(\frac{p_\phi^A}{R\sin\theta}\right)^2\right)+W(R,\theta,\phi)
\end{equation}
with
\begin{equation}
	p_R^A = p_R + A_R (R,\theta,\phi), \quad p_\theta^A = p_\theta + A_\theta (R,\theta,\phi), \quad p_\phi^A = p_\phi + A_\phi (R,\theta,\phi).
\end{equation}

\section{Generalized cylindrical case}\label{sec: ext cyl}
Let us now study the conditions for the existence of integrals of the form \eqref{int gen ext cyl} with $b=0$, namely
\begin{equation}\label{int ext cyl}
	X_1=(\ell^A_3)^2-a \ell^A_3 p^A_3+(c p^A_1 +d p^A_2) p^A_3+\ldots,\qquad X_2=(p^A_3)^2+\ldots
\end{equation}
Looking for generalized--type integrals, we assume that at least one of the real parameters $a,c,d$ is nonvanishing. (For results with $a=c=d=0$ see \cite{Fournier2019}.)
Because we assume $b=0$, we can simplify the integral $X_1$ further by rotating the system around the $z$ axis to set $d=0$. (This is equivalent to setting $d = C \sin(\cphi_0),\ c = C\cos(\cphi_0)$ with $\cphi_0=0$.)

Assumed integrability means the following conditions
\begin{equation}\label{integrability}
\{X_1,H\}=0, \quad \{X_2,H\}=0, \quad \{X_1,X_2\}=0,
\end{equation}
with the Hamiltonian of the form \eqref{HamMagnCyl}.

We will proceed in cylindrical coordinates $(r,\cphi,Z)$, where the full form of the integrals \eqref{int ext cyl} with $d=0$ is
\begin{equation}
	\begin{split} \label{cyl integrals}
		X_1={}&(p_\cphi^A)^2-a p^A_Z p^A_\cphi +\frac{c}{r} p^A_Z (r\cos\cphi p^A_r - \sin\cphi p^A_\cphi)
		+s_1^r p_r^A+s_1^\cphi p_\cphi^A+s_1^Z p_Z^A+m_1 , \\
		X_2={}&(p_Z^A)^2+s_2^r p_r^A+s_2^\cphi p_\cphi^A+s_2^Z p_Z^A+m_2,
	\end{split}
\end{equation}
the functions $s_i^{r,\cphi,Z}$ and $m_i\ (i=1,2)$ all depending on $(r,\cphi,Z)$.

Following the standard procedure, we equate to zero the coefficients of different monomials in $p_r,p_\cphi,p_Z$ of order $2,1$ and 0 in the conditions \eqref{integrability} (simplifying the lower order equations using those of higher order). We obtain from the second order terms 
\begin{equation}
	\begin{split}\label{cyl1sec}
		\partial_r s_1^r &=c \cos\cphi B^\cphi , \quad \partial_\cphi s_1^\cphi = -\frac{1}{r} s_1^r+\left( \frac{c \sin\varphi}{r}+a \right) B^r , \\
		\partial_\cphi s_1^r &=- r(c \sin\cphi+ a r) B^\cphi - c \cos\cphi B^r -r^2(\partial_r s_1^\cphi + 2 B^Z),\\
		\partial_\cphi s_1^Z &= c \cos\cphi B^Z -r^2(\partial_Z s_1^\cphi - 2 B^r), \quad 
		\partial_r s_1^Z = \left( \frac{c \sin\varphi}{r}+a \right) B^Z-\partial_Z s_1^r,
		\\ \partial_Z s_1^Z &= -c \cos\cphi B^\cphi -\left( \frac{c \sin\varphi}{r}+a \right) B^r,
	\end{split}
\end{equation}
and
\begin{equation}
	\begin{split}\label{cyl2sec}
		\partial_r s_2^r &= 0, \quad \partial_\cphi s_2^\cphi = -\frac{1}{r} s_2^r, \\
		\partial_\cphi s_2^r &= -r^2 \partial_r s_2^\cphi, \quad \partial_\cphi s_2^Z = -r^2 \partial_Z s_2^\cphi - 2 B^r, \\
		\partial_r s_2^Z &= -\partial_Z s_2^r + 2 B^\cphi, \quad \partial_Z s_2^Z = 0,
	\end{split}
\end{equation}
respectively.

The first order equations for $X_1$ reduce to
\begin{align} \label{cyl 1 fir}
	\partial_r m_1 &= s_1^Z B^\cphi - s_1^\cphi B^Z +c \cos\cphi \pd_Z W , \nonumber\\
	\partial_\cphi m_1 &=s_1^r B^Z - s_1^Z B^r + 2 r^2 \partial_\cphi W-(c \sin\cphi +a r) r \pd_Z W , \\
	\partial_Z m_1 &= s_1^\cphi B^r - s_1^r B^\cphi+c \cos\cphi\pd_r W -\left( \frac{c \sin\varphi}{r}+a \right) \pd_\cphi W, \nonumber
\end{align}
and for $X_2$ we have
\begin{align} \label{cyl 2 fir}
	\partial_r m_2 &= s_2^Z B^\cphi - s_2^\cphi B^Z, \nonumber\\
	\partial_\cphi m_2 &= s_2^r B^Z - s_2^Z B^r, \\
	\partial_Z m_2 &= s_2^\cphi B^r - s_2^r B^\cphi + 2 \partial_Z W \nonumber.
\end{align}
The zeroth order equations do not depend on the additional constants:
\begin{align} 
	&s_1^r \partial_r W + s_1^\cphi \partial_\cphi W + s_1^Z \partial_Z W = 0, \label{cyl1 0}\\
	&s_2^r \partial_r W + s_2^\cphi \partial_\cphi W + s_2^Z \partial_Z W = 0. \label{cyl2 0}
\end{align}

Imposing also involutivity of the integrals, i.e., the third equation in~\eqref{integrability}, we obtain further conditions. We write only the second order ones for now since we cannot easily eliminate their differential consequences in the lower order ones. We simplify the result by using the second order equations \eqref{cyl1sec} and \eqref{cyl2sec} to obtain
\begin{subequations}\label{extra2}
\begin{align}
\label{extra2.1}	&c \cos\cphi \pd_Zs^r_2&=0,\\
&(c \sin\cphi+ a r) \pd_Z s^\cphi_2+ 2 s^r_2&=0,\\
&2 r^2 \pd_r s^\cphi_2-c \cos\cphi \pd_Z s_2^\cphi+\left( \frac{c \sin\varphi}{r}+a \right)\pd_Z s_2^r&=0,\\
&r(c\sin \cphi+a r)\pd_r s^\cphi_2+c\sin\cphi s^\cphi_2 -2\pd_Z s^r_1&=0,\\
&2r^2\pd_Z s^\cphi_2-c \cos\cphi\pd_r s^\cphi_2 -\tfrac{1}{r} c \cos\cphi s_2^\cphi - \tfrac{1}{r}a s_2^r + 2 \pd_Z s^\cphi_1&=0,\\
&r(c\sin \cphi+a r)\pd_Z s^\cphi_2 -2\pd_Z s^Z_1&=0.
\end{align}
\end{subequations}

Second order equations for $X_2$ \eqref{cyl2sec} do not contain additional terms with constants $a,c,d$. We can therefore solve them for $s_2^{r,\cphi,Z},$ $B^\cphi$ and $B^r$ without much difficulty as in~\cite{Fournier2019}. Using the equations in \eqref{extra2} containing the derivatives of functions $s_1^{r,\cphi,Z}$ with respect to $Z$ and subsequently solving the fifth equation in \eqref{cyl1sec} for $B^Z$ (with our assumption of nonvanishing constants $a$ and $c$), all functions $s_i^{r,\cphi,Z},$ $i=1,2$ and $B^{r,\cphi,Z}$ are reduced as follows 
\begin{align}
	B^Z ={}& r D_{Z}S^\cphi_2(Z)+ \frac{r}{ c \sin\cphi +a r} \pd_R S^Z_1(r, \cphi),\nonumber\\
B^\cphi ={}& \tfrac{1}{2}\left(\pd_r S^Z_2(r, \cphi) + D^2_Z S^r_{21}(Z) \sin\cphi + D^2_Z S^r_{22}(Z) \cos\cphi\right),\nonumber\\
B^r ={}& \tfrac{1}{2}\left[-D^2_Z S^r_{21}(Z) r \cos\cphi +D^2_Z S^r_{22}(Z) r \sin\cphi - D^2_Z S^\cphi_2(Z) r^2 - \pd_\cphi S^Z_2(r, \cphi)\right],\nonumber\\
 s^Z_1 ={}& \tfrac{1}{2}\left[ a r \cos\cphi D_Z S^r_{21}(Z) - ( a r \sin\cphi +c) D_Z S^r_{22}(Z) +\right.\nonumber\\
{}&+\left.r(a r + c \sin\cphi) D_Z S^\cphi_2(Z) \right]+ S^Z_1(r, \cphi),\nonumber\\
 s^Z_2 ={}& S^Z_2(r, \cphi),\\
 s^\cphi_1 ={}& \frac{1}{2r} \left[c \cos\cphi S^\cphi_2(Z) + a \cos\cphi S^r_{22}(Z) + a \sin\cphi S^r_{21}(Z) \right] \nonumber\\
{}&- r^2 D_Z S^\cphi_2(Z) - r \cos\cphi D_Z S^r_{21}(Z) + r \sin\cphi D_Z S^r_{22}(Z)
 + S^\cphi_1(r, \cphi),\nonumber\\
 s^\cphi_2 ={}& \tfrac{1}{r}(D_Z S^r_{21}(Z)\cos\cphi - D_Z S^r_{22}(Z) \sin\cphi) +D_Z S^\cphi_2(Z),\nonumber\\
s^r_1 ={}&\tfrac{1}{2} \left[ - a \cos\cphi S^r_{21}(Z) + c \sin\cphi S^\cphi_2(Z) + a \sin\cphi S^r_{22}(Z)\right] + S^r_1(r, \cphi),\nonumber\\
 s^r_2 ={}& D_Z S^r_{21}(Z) \sin\cphi + D_Z S^r_{22}(Z) \cos\cphi.\nonumber
\end{align}
The remaining second order equations contain the constants $a,c$, so it is more convenient to split the considerations depending on their values. More concretely, the equation \eqref{extra2.1} reads
\begin{equation}\label{split eq 1}
	c (D^2_{Z} S^r_{22}(Z) \cos^2 \cphi +D^2_{Z} S^r_{21}(Z) \sin\cphi \cos\cphi)=0,
\end{equation}
so we have 2 major branches
\begin{enumerate}
	\item $c\neq0$, implies that 
\begin{equation}\label{split eq 1-1}
	S^r_{21}(Z) = b_3 c Z + S^r_{212}, \quad S^r_{22}(Z) = S^r_{221} Z + S^r_{222};
\end{equation}
	\item $c=0$, which implies $a\neq0$.
\end{enumerate}
We solve the first case in Subsection \ref{b0 cd nenula} and the second in Subsection \ref{b0 cd nula}.
\subsection{$c\neq0$}\label{b0 cd nenula}
By using \eqref{split eq 1-1}, the equations \eqref{extra2} can be almost solved by setting
\begin{equation}
S^\cphi_2(Z)=-S^\cphi_{22} Z^2 + S^\cphi_{21} Z + S^\cphi_{20},\quad S^r_{21}(Z) = S^\cphi_{22} c Z + S^r_{212},\quad S^r_{22}(Z) = S^r_{222},
\end{equation}
with the remaining equation
\begin{equation}
S^\cphi_{22} c a = 0,
\end{equation}
inducing another splitting.
\begin{enumerate}
\item Assuming that $S^\cphi_{22}\neq0$ (and $a=0$), we can solve for $S^Z_2$ from the fourth and subsequently the last equation in \eqref{cyl1sec}. The first and second equations in \eqref{cyl1sec} then yield $S^\cphi_1.$ The remaining second order equations further constrain these functions. 

We proceed with Clairaut's compatibility conditions
\begin{equation}
	\pd_i \pd_j m_k= \pd_j \pd_i m_k,\qquad i,j\in\{r,\cphi,Z\}, \quad k=1,2,
\end{equation}
where we substitute for $m_k$ from \eqref{cyl 2 fir} or \eqref{cyl 1 fir}. Those for $m_2$ partially determine the form of $W$, which we further constrain by \eqref{cyl2 0}. With these results, the compatibility conditions for $m_1$ can be solved and we arrive at the magnetic field $B$ as given below. The remaining equations coming from $\{X_1,H\}=0$, namely~\eqref{cyl 1 fir} and~\eqref{cyl1 0}, further constrain $W$. The complete solution is described by the following magnetic field $B$ and  potential\footnote{For better readability, we have renamed all the integration constants arising in the computation, e.g. $S^\cphi_{2j},S^r_{2j2}$, so that $b_i$ are present in the magnetic field, $w_j$ are only in the potential $W$; the same applies in the other cases below.} $W$

\begin{align}
	B^r ={}& r \left(b_3 r+c b_2 \cos\cphi \right),\quad B^\cphi = - c b_2 \sin\cphi,\quad 	B^Z = -(2 b_3 Z +3 b_2 r^2+ b_1) r,\notag\\
	W={}& \frac{b_3^2}{8} Z^4 - \frac{b_3 w_1}{2}Z^3 - Z^2 \left(\frac{b_3^2}{2}r^2-\frac{ b_2 b_3}{2} c r \cos\cphi +\frac{w_2b_3}{4} - \frac{w_1^2}{2}\right) \notag \\
	{}&- Z \left(\frac{3 b_2 b_3 }{4}r^4+\frac{b_3 b_1}{2} r^2 - \frac{b_3^2-2 b_2 w_1}{2} c r \cos\cphi +\frac{w_3b_3}{2} - \frac{w_2 w_1}{2}\right) \label{system c<>0}\\
	{}&- \frac{b_2^2}{4}r^6+\frac{(w_1 - b_1) b_2} {4}r^4 +\frac{ b_2 b_3}{2} c r^3\cos\cphi +\frac{b_2^2}{2} c^2 r^2 \cos^2\cphi \notag\\
	{}&+ \left( \frac{w_1 (b_1+ w_1)}{2} - \frac{b_2 w_3}{2}\right) r^2+ \frac{ b_3 (b_1 + w_1) - b_2 w_2}{2} c r \cos\cphi.\notag
\end{align}
Transforming into Cartesian coordinates, we have
\begin{align}
	B^x ={}& b_3 x+ c b_2 ,\quad B^y = b_3 y,\quad B^z = -(2 b_3 z +3 b_2 r^2+b_1),\notag\\
		W={}& \frac{b_3^2}{8} z^4 - \frac{b_3 w_1}{2}z^3 - z^2 \left(\frac{b_3^2}{2}r^2-\frac{ b_2 b_3}{2} c x +\frac{w_2b_3}{4} - \frac{w_1^2}{2}\right) \notag\\
	{}&- z \left(\frac{3 b_2 b_3 }{4}r^4+\frac{b_3 b_1}{2} r^2 - \frac{b_3^2-2 b_2 w_1}{2} c x +\frac{w_3b_3}{2} - \frac{w_2 w_1}{2}\right) \\
	{}&- \frac{b_2^2}{4}r^6+\frac{(w_1 - b_1) b_2} {4	}r^4 +\frac{ b_2 b_3}{2} c x r^2
	+ \frac{b_2^2}{2} c^2 x^2 +\notag\\
	{}&+\left(\frac{w_1 (b_1+ w_1)}{2}- \frac{b_2 w_3}{2}\right) r^2+ \frac{ b_3 (b_1 + w_1) - b_2 w_2}{2} c x,\notag
\end{align}
with the shorthand notation $r^2=(x^2+y^2)$.

In the preceding calculation we also obtained the form of the functions in integrals \eqref{cyl integrals} which we hereby list:
\begin{align}
	s^r_1={}& -\frac{c b_3}{2}Z^2 \sin \cphi+ c w_1 Z \sin \cphi - c^2b_2 r\sin \cphi \cos \cphi + \frac{c w_2}{2}\sin \cphi,\notag \\
	s^\cphi_1={}& -\frac{c b_3}{2r}Z^2 \cos \cphi + Z \left(2 b_3 r^2+\frac{c w_1 }{r}\cos \cphi\right) + \frac{3b_2}{2} r^4-c b_3 r \cos \cphi \notag \\
	{}&-c^2 b_2 \cos^2\cphi + b_1 r^2+ \frac{c w_2}{2r}\cos \cphi + w_3,\label{X1 prvni}\\
\notag		s^Z_1={}& -c b_3 r Z \sin \cphi - c b_2 r^3 \sin \cphi - c (b_1+ w_1) r \sin \cphi, 
		\end{align}
\begin{align*}
m_1={}& -\frac{c b_3^2}{2}rZ^3\cos \cphi+ Z^2\left(b_3^2 r^4 - \frac{b_2 b_3 c}{2}r^3 \cos \cphi- \frac{c b_3 (b_1- w_1) }{2} r\cos \cphi+ \frac{c^2b_3^2}{4}\right)\notag \\
{}&+ \left[\frac{3 b_3 b_2}{2}r^6+ b_3 b_1 r^4 + (b_2 w_1- b_3^2) c r^3 \cos \cphi - b_3 r^2\left(c^2 b_2\cos^2\cphi - \frac{c^2 b_2}{2}- w_3\right)\right.\notag \\
{}&+\left.\frac{ c (w_2b_3 + 2b_1w_1+ 2w_1^2)}{2} rZ \cos \cphi + \frac{c^2b_3 (b_1+ w_1)}{2}\right] \notag \\
 {}&+ \frac{9 b_2^2}{16}r^8 + \frac{3 b_2b_1}{4}r^6 -\frac{3 c b_2 b_3 }{4} r^5 \cos \cphi\notag \\
 {}&- \left(c^2 b_2^2\cos^2\cphi-\frac{c^2 b_2^2}{4} - \frac{3 b_2 w_3}{4} - \frac{b_1^2}{4}\right) r^4 +\frac{c (w_2 b_2- b_3 b_1) }{2} r^3\cos \cphi\notag \\
 {}&+ r^2\left[\frac{c^2(b_3^2-4(b_1+ w_1) b_2)}{4} \cos^2\cphi +\frac{c^2 b_2(b_1+ w_1) }{2}+ \frac{b_1 w_3}{2}\right] \notag \\
 {}&+ \frac{c}{2} \left[(b_1+ w_1) w_2- b_3 w_3)\right] r\cos \cphi ,\notag
\end{align*}
\begin{align}
	s^r_2={}&c b_3 \sin \cphi,\quad s^\cphi_2= -2 b_3 Z + 2 w_1+ \frac{c b_3}{r}\cos \cphi,\quad s^Z_2=-2c b_2 r\sin \cphi,\label{X2 prvni} \\
	m_2={}& \frac{b_3^2}{4}Z^4 - b_3 w_1 Z^3-Z^2 (2b_3^2r^2+ \frac{w_2b_3}{2} - w_1^2) \notag \\
	{}&+Z \left[-\frac{3 b_2b_3}{2}r^4- b_3 (b_1- 2w_1) r^2+ 2c b_3^2 r \cos \cphi+ (c^2 b_2- w_3) b_3 + w_1 w_2\right] \notag \\
	{}&+ \frac{3 b_2 w_1}{2}r^4+ c b_2 b_3 r^3\cos \cphi + r^2 (c^2b_2^2\sin^2\cphi + b_1w_1)+ c b_3 b_1 r\cos \cphi. \notag
\end{align}

We notice that the solution~\eqref{system c<>0} with the corresponding integrals exists also for $S^\cphi_{22}\equiv b_3=0$, cf.~\eqref{system c<>0 red} below.

A straightforward, however tedious, computation shows that no additional quadratic integrals of motion exist for the system of the form~\eqref{system c<>0} unless $b_3=b_2=0$. If that is the case, the magnetic field is constant, and the system may admit an additional second order integral only if $W=W(r)$. However, under this assumption the system admits two cylindrical--type integrals which become $Y_1=p_\cphi=\ell_z$, $Y_2=p_Z=p_z$ in a suitable choice of gauge. This makes the constant $a$ in $X_1$ irrelevant since both $\ell_z^2$ and $\ell_z p_z$ are themselves integrals. The system separates in cylindrical coordinates and has therefore been analyzed in \cite{Kubu_2021}.

\item $S^\cphi_{22}=0$: It is too complicated to solve \eqref{cyl1sec} at this moment, so we turn to the first order equations. Having solved \eqref{extra2}, we can now simplify the first order equations from $\{X_1,X_2\}$. These equations together with \eqref{cyl 1 fir} and \eqref{cyl 2 fir} lead to 
\begin{equation}
	m_1 = M_{11}(r, \cphi) Z + M_{12}(r, \cphi),\quad m_2 = \tfrac{1}{4} (S^\cphi_{21})^2 Z^2 + m_{20} Z+\tfrac{1}{4} S^Z_2(r, \cphi)^2+ M_{21}(r),
\end{equation}
where $m_{20}$ is a constant and $M_{ij}$ are yet undetermined functions of the respective variables. Furthermore, we obtain
\begin{equation}
	W = \tfrac{1}{8}(S^\cphi_{21})^2 Z^2 + \left(\tfrac{m_{20}}{2} + \tfrac{1}{4} S^\cphi_{21} \pd_\cphi S^Z_2(r, \cphi) \right) Z + W_1(r, \cphi),
\end{equation}
with $W_1$ yet another undetermined function, and provide the following splitting term
\begin{equation}
	a (S^\cphi_{21})^2=0.
\end{equation}

\begin{enumerate}
	\item If $S^\cphi_{21}\neq0$, equation \eqref{cyl2 0} simplifies $S^Z_2$ to
	\begin{equation}
		S^Z_2 = B^\cphi_1(r)\sin\cphi + B^\cphi_2(r)\cos\cphi.
	\end{equation}
	The first order equations from $\{X_1,X_2\}$ can be solved for $S^\cphi_1$. The last equation in \eqref{cyl 2 fir} determines $B^\cphi_1$ in terms of $B^\cphi_2$ and we can therefore solve for both from \eqref{cyl1 0}, the result being
	$B^\cphi_1(r)=-2 c b_2 r,\ B^\cphi_2(r)=0.$
	All the remaining equations can now be solved (first \eqref{cyl1sec}, next \eqref{cyl1 0} followed by the equations coming from $\{X_1,X_2\}=0$ and then the rest).
	
The solution for the magnetic field $B$ and potential $W$ reads 
\begin{align}
	B^r ={}& c b_2 r \cos\cphi,\quad B^\cphi = - c b_2 \sin\cphi,\quad 	B^Z = -(3 b_2 r^2+ b_1) r,\notag\\
	W={}& \frac{w_1^2}{2} Z^2 -Z \left(c b_2 w_1 r \cos\cphi- \frac{w_2 w_1}{2} \right)- \frac{b_2^2}{4}r^6+\frac{(w_1 - b_1) b_2} {4}r^4 \label{system c<>0 red}\\
	{}&+\frac{b_2^2}{2} c^2 r^2 \cos^2\cphi+ \left( \frac{w_1 (b_1+ w_1)}{2}- \frac{b_2 w_3}{2}\right) r^2-\frac{b_2 w_2}{2} c r \cos\cphi.\notag
\end{align}
This corresponds to the case in \eqref{system c<>0} with $b_3=0$, so there is no superintegrable case unless $b_2=0$ as well as $W=W(r)$, cf. discussion below equation \eqref{X2 prvni}. The form of the integrals $X_1$ and $X_2$ can be obtained from \eqref{X1 prvni} and \eqref{X2 prvni} by setting $b_3=0$.

\item If $S^\cphi_{21}=0$, equations \eqref{cyl 2 fir} impose $M_{21}=0$ and the first order equations from $\{X_1,X_2\}$ eliminate all dependence on $Z$. This implies that the integral $X_2$ reduces to the first order integral $\tilde{X}_2=p_Z$ in a suitable gauge as the coordinate $Z$ is cyclic. 

At this moment, only equations coming from $\{X_1,H\}=0$ remain, all the others are solved. Using the last equation in \eqref{cyl1sec}, $S^Z_2$ is reduced to a function of one variable
\begin{equation}
	S^Z_2 = S^Z_{21}\left(\tfrac{1}{2}a r^2 + c r\sin\cphi\right),
\end{equation}
and subsequently $S^Z_1$ is found from the fourth equation as well,
\begin{equation}
	S^Z_1= r^3(D S^Z_{21})(\tfrac{1}{2}a r^2 + c r\sin\cphi) \frac{4 c \sin\cphi + 3 a r}{8}
	+S^Z_{11}(\tfrac{1}{2}a r^2 + c r\sin\cphi),
\end{equation}
where $D S^Z_{21}$ denotes $S^Z_{21}$ differentiated once with respect to its argument $\tfrac{1}{2}a r^2 + c r\sin\cphi$. The first equation in \eqref{cyl1sec} then yields $S^\cphi_1$ in a form of an unevaluated integral of the functions defined above. Using this in the third equation \eqref{cyl1sec} and differentiating with respect to $r$ and $\cphi$ to eliminate the unevaluated quadrature, we obtain an equation with functions $S^Z_{11}$ and $S^Z_{21}$ only. To deal with this equation, we define
 \begin{equation}
 	\xi=\tfrac{1}{2}a r^2 + c r\sin\cphi,
 \end{equation}
solve for $\cphi$ and eliminate it from the equation. We find
\begin{equation}\label{eq xi}
\begin{split}
		&D^4 S^Z_{21}(\xi) a^2 r^6 -6 a(D^4 S^Z_{21}(\xi) \xi +6 D^3 S^Z_{21}(\xi)) r^4&\\
		&+ 8(D^4 S^Z_{21}(\xi) \xi^2 + 11 \xi D^3 S^Z_{21}(\xi) - a D^3 S^Z_{11}(\xi) + 24 D^2 S^Z_{21}(\xi)) r^2&\\
		& + 16 \xi D^3 S^Z_{11}(\xi) + 64 D^2 S^Z_{11}(\xi)&=0,
\end{split}
\end{equation}
where all derivatives (denoted by $D$) are with respect to $\xi$. Since there is no functional dependence on $\xi$, the coefficients of each power of $r$ must vanish in~\eqref{eq xi}. The solution depends on the value of $a$.

When $a\neq0$ we find
\begin{equation}
	 S^Z_{11}=2 b_1 \xi + b_3,\quad S^Z_{21}=2 b_2 \xi+b_4,
\end{equation}
which determines the magnetic field $B$. Finally, equations \eqref{cyl 1 fir} and \eqref{cyl1 0} give the potential $W$, the result reading
\begin{equation}\label{xi system}
\begin{split}
		B^r ={}& b_2 c r \cos\cphi,\quad B^\cphi=-b_2 (a r +c \sin{\cphi}),\quad B^Z = -(3 b_2 r^2+2 b_1) r,\\
		 W={}&b_2\left[-\frac{b_2}{4} r^6 - \frac{ b_2 a^2+4 b_1}{8}r^4 - \frac{a c b_2}{2}r^3 \sin\cphi \right.\\
		 {}&+ \left.r^2\left(\frac{b_2 c^2}{2} \cos^2\cphi+w_1\right) +r( w_2 \cos\cphi +w_3 \sin\cphi)\right].
\end{split}
\end{equation}	

In Cartesian coordinates, we have
\begin{equation}
\begin{split}
	B^x={}& b_2 (a y+c),\quad B^y=-b_2 a x,\quad B^z=-(3 b_2r^2+2 b_1),\\
	W ={}& b_2\left[-\frac{b_2}{4} r^6 - \frac{ b_2 a^2+4 b_1}{8}r^4 - \frac{a c b_2}{2}r^2 y + \frac{b_2 c^2}{2} x^2+w_1 r^2 +w_2 x +w_3 y\right],
\end{split}
\end{equation}
where we write $r^2=x^2+y^2$ for brevity. If $b_2$ vanishes, we obtain the constant magnetic field again.

Let us see the form of the integrals. As we have already noted, $Z$ is a cyclic coordinate in a suitable choice of gauge and therefore $\tilde{X}_2=p_Z^A+m_2$, where $m_2=-A_Z$, is a reduced form of $X_2$. The form of $X_1$ can be determined by
\begin{align}
	s^Z_1 ={}& -c b_2 r^3\sin \cphi - 2 c b_1 r \sin \cphi - \frac{3 a b_2}{4} r^4 - a b_1 r^2,\notag\\
	s^\cphi_1 ={}& \frac{3 b_2}{2} r^4 + \frac{b_2 a^2 + 4 b_1}{2} r^2 + \frac{3 a c b_2}{2}r\sin \cphi -c^2 b_2\cos^2\cphi \notag\\
	{}&- 2 w_1 - \frac{w_2\cos \cphi+w_3 \sin \cphi }{r},\notag\\
	s^r_1 ={}& -\frac{a c b_2}{2} r^2 \cos \cphi - c^2 b_2 r \sin \cphi \cos \cphi - w_2 \sin \cphi + w_3 \cos \cphi,\label{treti}\end{align}
	\begin{align*}
	m_1 ={}& \frac{9 b_2^2}{16} r^8 + \frac{3 b_2 (a^2 b_2 + 4 b_1)}{8} r^6 + \frac{5 a c b_2^2}{4} r^5 \sin \cphi \notag\\
	{}&+r^4\left[-c^2 b_2^2\cos \cphi^2 + \frac{c^2 b_2^2}{4} + \frac{(a^2 b_1 - 3 w_1) b_2}{2} + b_1^2\right] \notag\\
	{}&+ b_2 r^3 [(2 a c b_1 - w_3)\sin \cphi - w_2 \cos \cphi] \notag \\
	{}& - r^2 (2\cos \cphi^2 b_2 c^2 - c^2 b_2 +2 w_1) b_1 - 2 b_1 r (\sin \cphi w_3 + \cos \cphi w_2). \notag
		\end{align*}
Searching for superintegrability we find that the system does not admit any additional integral of motion at most second order in momenta.\medskip

If $a=0,$ the solution to \eqref{eq xi} contains more integration constants $b_j$
\begin{equation}
		 S^Z_{11}=b_1 \xi + b_3+4 b_5 \xi^{-2},\quad S^Z_{21}=2 b_2 \xi+b_4+b_6 \xi^{-2}+b_7 \xi^{-4}.
\end{equation}
However, using the remaining equations, many of the constants $b_1,\ldots,b_7$ must vanish and	we obtain the following two systems, neither of which new. The first one is given by
		\begin{align}
\begin{split}
		B^r&={\frac{4 b_5 \cos\cphi}{c^2 r^2 \sin^3 \cphi}},\quad 	{B^\cphi} =-{\frac{4 b_5}{c^2 r^3 \sin^2 \cphi}},\quad {B^Z} =0,\\
		W &=-4\left(\frac{b_5^2} {2c^4 r^4 \sin^4 \cphi} +\frac{W_0}{c^2 r^2 \sin^2 \cphi}\right), \raisetag{10pt}
\end{split}
	\end{align}
which in Cartesian coordinates become
		\begin{align}\label{WB netriv}
			B^x={\frac{4 b_5}{c^2 y^3}},	\quad {B^y} =0,\quad B^z=0,\quad 	W =-4\left(\frac{b_5^2}{2 c^4 y^4}	+\frac{W_0 }{c^2 y^2}\right).
		\end{align}
	This is a superintegrable system at the intersection of Cases Ia and Ib in~\cite{Marchesiello2019}, i.e., it admits two first order integrals $p_x,p_z$ (in a suitably chosen gauge) and $X_1$ separates into two integrals $\ell_z^2+\ldots$ and $\ell_z p_z+\ldots$, which makes the constant $a$ irrelevant. It is separable in Cartesian coordinates (hence integrable with Cartesian--type integrals) and moreover integrable with cylindrical--type integrals~\cite{Kubu2020}.
 
The second system we find is a special case of \eqref{xi system} with $a=0,$ namely
\begin{align}
		B^r & = b_2 c r \cos\cphi,\quad B^\cphi=-b_2 c \sin{\cphi},\quad B^Z = -(3 b_2 r^2+b_1) r,\\ \nonumber
		W & =-b_2\left[\frac{b_2}{4} r^6 + \frac{b_1}{2}r^4 -r^2 \left(\frac{b_2 c^2}{2} \cos^2\cphi+w_1\right) 
		-r( w_2 \cos\cphi +w_3 \sin\cphi)\right].
\end{align}
The form of the integral $X_1$ is obtained by setting $a=0$ in \eqref{treti}.
\end{enumerate}
\end{enumerate}
To summarize the case $c\neq 0$, we have found two new integrable systems, namely \eqref{system c<>0} and \eqref{xi system}.

\subsection{$c=0$ and $a\neq0$}\label{b0 cd nula}
In this case, all second order equations, i.e., \eqref{cyl1sec}, \eqref{cyl2sec} and \eqref{extra2}, can be solved. From \eqref{extra2} we obtain
\begin{equation}
\begin{split}
		S^r_{21}(Z) ={}& \frac{w_2-2s_1^r}{a} -\frac{b_3 a^2}{2} \cos\left(\frac{2 Z}{a}+Z_0\right),\quad S^r_{22}(Z) = \frac{w_3+2s_2^r}{a}-\frac{b_3 a^2}{2} \sin\left(\frac{2 Z}{a}+Z_0\right),\\
		S^\cphi_2(Z) ={}& s^\cphi_{21} Z + s^\cphi_{22},\raisetag{\baselineskip}
\end{split}
\end{equation}
where $b_3,Z_0$ are constants as well as lower case $s_i,w_i$ (notation is chosen for later convenience; $b_3,w_2,w_3$ figure in the final expression for the magnetic field and potential). Because the translation $\tilde{Z}=Z+Z_0$ does not change the momenta and we can redefine the functions depending on $Z$, we can assume that $Z_0=0$ without loss of generality.

Now equations \eqref{cyl1sec} imply
\begin{equation}
\begin{split}
		S^\cphi_1(r, \cphi) ={}& \tfrac{1}{r}(s^r_1 \cos\cphi - s^r_2\sin\cphi ) - \tfrac{a}{2} S^Z_{21}(r) - \tfrac{2}{a} S^Z_{11}(r) + s^\cphi_{11}-\tfrac{1}{4}s^\cphi_{21}a^2,\\
S^r_1(r, \cphi) ={}& s^r_1\sin\cphi + s^r_2\cos\cphi ,\quad S^Z_1(r, \cphi) = S^Z_{11}(r),\quad 	 S^Z_2(r, \cphi) = S^Z_{21}(r),
\end{split}
\end{equation}
where lower case $s_i$ are constants and $S_i$ are yet undetermined functions of the indicated variable.

We can now derive the first order equations from $\{X_1,X_2\}$. By equating to zero the coefficient of $p_r$ in particular, we find
\begin{equation}
	\begin{split}
	s^\cphi_{21} [ w_2\cos\cphi +w_3 \sin\cphi ]	-(s^\cphi_{21} a^2 + 4 s^\cphi_{11})\frac{b_3 a}{2} \sin\left(\cphi +\frac{ 2 Z}{a}\right)=0,
	\end{split}
\end{equation}
which leads to the second splitting.
\begin{enumerate}
	\item $b_3\neq 0$: We solve \eqref{cyl 2 fir} for $m_2$ and $W$; moreover, $S^Z_{21}$ can be determined from $S^Z_{11}$ up to a constant. The first order equations from $\{X_1,X_2\}$ further constrain these functions and partially determine $m_1$. We similarly use \eqref{cyl 1 fir} as further constraints on $m_i$ and $W$. Equation \eqref{cyl2 0} then determines $S^Z_{11}$ (and also $S^Z_{21}$), yielding our final magnetic field $B$, the last unknown function in $W$ and eliminates some constants. The remaining function in $m_1$ is determined through \eqref{cyl 1 fir}. 
	
The final result for the magnetic field $B$ and potential $W$ is
\begin{align}
		B^r ={}& -b_3 r \cos\left(\cphi +\frac{ 2 Z}{a}\right),\	B^\cphi= b_2 r +b_3 \sin\left(\cphi +\frac{ 2 Z}{a}\right),\ B^Z= r\left(\frac{b_2}{a} r^2 +b_1\right),\notag\\		
W={}&-(b_2 a^2 -w_1) \left[\frac{b_2}{8 a^2} r^4 + \frac{2b_1 a + w_1}{8 a^2} r^2+
\frac{b_3}{4} r\sin\left(\cphi +\frac{ 2 Z}{a}\right)\right].\label{solutionc=0b3<>0}
\end{align}
Transforming into Cartesian coordinates, the result reads
\begin{equation}
	\begin{split}
	B^x={}&-b_2 y-b_3\cos\left(\frac{2z}{a}\right),\quad B^y=b_2 x+b_3\sin\left(\frac{2z}{a}\right),\quad B^z=\frac{ b_2}{a}r^2+b_1,\\
	W={}&-(b_2 a^2 -w_1) \left[\frac{b_2}{8 a^2} r^4 + \frac{2 b_1 a + w_1}{8 a^2} r^2+ \frac{b_3}{4} \left(x \sin\left(\frac{2z}{a}\right) + y \cos\left(\frac{2z}{a}\right)\right)\right].
\end{split}
\end{equation}
For the integral $X_2$ we find
\begin{equation}
\begin{split}
			s^Z_2 = {}&b_2 r^2 + a b_1 + \frac{w_1}{2},\quad s^\cphi_2 = \frac{b_3 a}{r}\sin\left(\cphi +\frac{ 2 Z}{a}\right),\quad s^r_2 = -b_3 a\cos\left(\cphi +\frac{ 2 Z}{a}\right),\\
m_2={}&\frac{b_2^2}{4}r^4 + \frac{b_2 (2 a b_1 + w_1)}{4}r^2 + \frac{b_3 w_1}{2}r \sin\left(\cphi +\frac{ 2 Z}{a}\right).
\end{split}
\end{equation}
This system admits a first order integral
\begin{equation}
	Y_3=p_\cphi^A-\frac{a}{2}p_Z^A-\frac{b_2}{4a}r^4 -  \frac{1}{4} \left( a b_2 + 2 b_1 \right) r^2 - \frac{a b_3}{2} r\sin\left(\cphi +\frac{ 2 Z}{a}\right)
\end{equation}
and the integral $X_1$ is dependent on $Y_3$ and $X_2$, namely
\begin{equation}
	X_1=Y_3^2-\frac{a^2}{4} X_2;
\end{equation}
thus, $Y_3$ can be seen as a square root of $X_1$ in this sense.

Solving the determining equations \eqref{second QM cart}--\eqref{zeroth QM cart} with the magnetic field and potential~\eqref{solutionc=0b3<>0} for hypothetical additional integral, we find that in this class only the system with $b_2=0$ and $W=0$ is quadratically superintegrable, with three first order integrals, see Section \ref{nonsep Cart} below for further analysis.
\item $b_3=0$: Equations \eqref{cyl 2 fir} reduce the form of $m_2$ and $W$, subsequently \eqref{cyl 1 fir} partially determine $m_1$ and simplify $W$ further. We obtain the following equation yielding yet another split
\begin{equation}\label{spl}
\begin{split}
		[w_2 \cos\cphi+w_3 \sin\cphi ] (D^2_r S^Z_{21}(r) r^3 + 2 D_r S^Z_{21}(r) r^2 - D^2_r S^Z_{11}(r) r +D_r S^Z_{11}(r))=0.
\end{split}
\end{equation}

In the first case, where the square bracket in \eqref{spl} vanishes and $S^Z_{11}(r)$ remains independent of $S^Z_{21}(r)$, equations \eqref{cyl 1 fir} completely determine $m_1$. Then all equations are solved with the solution
\begin{align}
		\begin{split}
		B^r ={}&0,\quad 		B^\cphi= \tfrac{1}{2}\pd_r S^Z_{21}(r),\quad 
		B^Z=s^\cphi_{21} r+\frac{1}{a} \pd_r S^Z_{11}(r),\quad 	
		W=W(r).
	\end{split}
\end{align}
It is clear that this system admits two cylindrical integrals $Y_1=p_\cphi=\ell_z$, $Y_2=p_Z=p_z$, implying that $X_2=Y_2^2$ reduces to a first order integral and we can separate the integral $X_1$ into integrals $\ell_z p_z+\ldots$ and $\ell_z^2+\ldots$, making the constant $a$ irrelevant. The system thus separates in cylindrical coordinates and its further analysis can be found in \cite{Kubu_2021}.

In the second case, where the square bracket in \eqref{spl} does not vanish, the functions $S^Z_{11}$ and $S^Z_{21}$ are related by it. Excluding the case with constant magnetic field, equation \eqref{cyl2 0} implies $s^\cphi_{21}=0$. Equation \eqref{cyl1 0} then determines $S^Z_{21}$ and the remaining functional dependence in $W$. $S^Z_{11}$ is subsequently determined through \eqref{spl} and we obtain the result
\begin{equation}\label{druhy}
	\begin{split}
	B^r ={}&0,\quad B^\cphi=-b_2 a r,\quad B^Z = -(3 b_2 r^2+2b_1) r,\\
W={}&b_2\left[-\frac{b_2}{4} r^6 - \frac{ b_2 a^2+4 b_1}{8}r^4 +w_1 r^2 +r( w_2 \cos\cphi +w_3 \sin\cphi)\right].
	\end{split}
\end{equation}
This system also admits $p_Z$ as an integral because the coordinate $Z$ is cyclic (in a suitable gauge), and it is a special case of \eqref{xi system} with $c=0$. The form of the integral $X_1$ is obtained by setting $c=0$ in \eqref{treti}.

Superintegrability calculations assuming $B^\cphi=B^\cphi(r),B^Z=B^Z(r)$ and arbitrary $W$ (i.e., both cases at once) show that only the cylindrical case with $W=W(r)$ and two first order integrals $Y_1=p_\cphi=\ell_z$, $Y_2=p_Z=p_z$ admits an additional at most second order integral. 
 \end{enumerate}

To summarize the case $c=0$, $a\neq 0$, the only new system found here is \eqref{solutionc=0b3<>0}. When $b_2=0$, this system also admits Cartesian--type integrals, thus it is even superintegrable, though not separable, as we will show in the next section.

\section{A quadratically superintegrable but not separable system}\label{nonsep Cart}

In the generalized cylindrical case, we found the following superintegrable Cartesian system: The magnetic field $\vec{B}$, potential $W$ and the vector potential $\vec{A}$ in our chosen gauge are
\begin{equation}\label{super3int1radu}
	\begin{split}
		\vec{B}\left(x, y, z\right) ={}&\left( -b_3 \cos\left(\frac{2z}{a}\right),b_3 \sin\left(\frac{2z}{a}\right), b_1\right),\quad W(x,y,z)=0,\\
		\vec A\left(x,y,z \right) ={}&\left(-\frac{b_3 a}{2} \cos\left(\frac{2z}{a}\right), b_1 x + \frac{b_3 a}{2}\sin\left(\frac{2z}{a}\right),0\right).
	\end{split}
\end{equation}

If $b_3=0$, the system reduces to a well--known superintegrable system with constant magnetic field, which has been analyzed in \cite{Landau}, see also \cite{Marchesiello2015}. The superintegrability of the system with $b_1=0$ is known as well, see \cite{Marchesiello2015}. As observed in~\cite{HeiIld}, it describes motion of electrons in a nonrelativistic limit of a helical undulator. (Undulators are magnetic devices for generation of powerful coherent radiation using beams of charged high energy particles, typically electrons.)

We therefore continue the analysis assuming both $b_3\neq0$ and $b_1\neq0$. Thus, physically we study the motion of (nonrelativistic) electrons in the field of a helical undulator placed in an infinite solenoid, which was recently proposed as a simple and efficient source of coherent spontaneous THz undulator radiation, see~\cite{PhysRevAccelBeams.20.122401}.

The system admits three first order integrals reading
\begin{align}
	Y_1 ={}& p_x^A + b_1 y + \frac{b_3 a}{2} \cos\left(\frac{2z}{a}\right)\label{SI Y1},\\
	Y_2 ={}& p_y^A - b_1 x - \frac{b_3 a}{2} \sin\left(\frac{2z}{a}\right),\label{SI Y2}\\
	\begin{split}
		Y_3 ={}& \ell_z^A-\frac{a}{2} p_z^A -\frac{1}{2}\left[b_1 r^2 + b_3 a\sin\left(\frac{2z}{a}\right) x+ b_3 a \cos\left(\frac{2z}{a}\right) y\right].\label{SI Y3}
	\end{split}
\end{align}
These first order integrals do not commute, as their Poisson brackets read
\begin{equation}\label{SI alg}
	\{Y_1,Y_2\}=b_1,\quad \{Y_1,Y_3\}=-Y_2,\quad \{Y_2,Y_3\}=Y_1.
\end{equation}
These integrals constitute the solvable Lie algebra $\mathfrak{s}_{4,7}\oplus \mathfrak{a}_1$ (where $\mathfrak{s}_{4,7}$ is the 7th solvable 4D algebra in \cite{Snobl_book}), the (canonical) basis for $\mathfrak{s}_{4,7}$ being $-b_1 I,Y_2,Y_1,Y_3$, where the identity $I$ commutes with all $Y_i$, and the Hamiltonian $H$ constitutes the one--dimensional abelian Lie algebra $\mathfrak{a}_1$. The Casimir invariants are $H, I$ and
\begin{equation}
	K=Y_1^2+Y_2^2+2 b_1 Y_3.
\end{equation}
By an explicit calculation we find that all second order integrals are functions of integrals $H,Y_i$. Consequently, the system~\eqref{super3int1radu} has the same form in quantum mechanics as well because for first order integrals no quantum corrections arise \cite{Marchesiello2015}.

The Casimir invariant $K$ commutes with all integrals of order at most 2, as well as the Hamiltonian. Thus, instead of $K$ we may equivalently consider our integral $X_2=2 H - K$:
\begin{equation}\label{SI X2}
	\begin{split}
		X_2={} \left(p_z^A\right)^2 - b_3 a \cos\left(\frac{2z}{a}\right) p_x^A + b_3 a \sin\left(\frac{2z}{a}\right) p_y^A + a b_1 p_z^A - \frac{b_3^2 a^2}{4}.
	\end{split}
\end{equation}
Every commuting triple of quadratic integrals can then be written as a linear span of $H, X_2$ and another second order integral constructed out of $Y_1,Y_2,Y_3$. Its role can be played by, e.g., the integral $X_1$ of~\eqref{int ext cyl} which can be expressed as
\begin{equation}
	X_1=Y_3^2-\frac{a^2}{4} X_2=Y_3^2+\frac{a^2}{4}(Y_1^2+Y_2^2+2 b_1 Y_3-2 H),
\end{equation}
or by the integral $Y_1^2$; thus, the system~\eqref{super3int1radu} can be written as an integrable system of the Cartesian type and identified as Case 5 in \cite{Zhalij_2015} (with permuted coordinates).

Matching the possible choices of commuting triples of quadratic integrals of the system~\eqref{super3int1radu} with the general structure of~\cite{Marchesiello2022}, we see that the system~\eqref{super3int1radu} lies at the intersection of classes (f) and (k) of Theorem 1 therein and does not belong to any other class. 

Unlike the version with $b_3=0$ from \cite{Marchesiello2015}, this system is not separable in Cartesian coordinates, for one of the Levi-Civita separation conditions \cite{LeviCivita1904} (no sum over indices $i\neq j$, $\pd_i\equiv\pd_{x^i},\pd^j\equiv \pd_{p_j}$)
\begin{equation}
	\pd^i\pd^jH \pd_i H\pd_j H+\pd_i\pd_j H \pd^i H \pd^j H-\pd^i\pd_j H \pd_i H\pd^j H- \pd_i\pd^j \pd^i H \pd_j H=0
\end{equation}
does not vanish as $b_1 b_3\neq0.$

We therefore conclude that we have found a minimally superintegrable system, with Cartesian--type integrals, which does not separate in Cartesian coordinates. Neither it is separable in any other orthogonal coordinate system connected with the separation of the Hamilton--Jacobi equation, for the following reason. The results of \cite{Benenti_2001} imply that also in the presence of magnetic field, the separation relates to at most quadratic integrals, at least one of them of order one. Equation~(6.11) in \cite{Benenti_2001} also determines that the corresponding quadratic integral differs from the purely scalar case by the presence of linear terms in momenta only. Consequently, the separation can occur only in the original set of 11 orthogonal coordinate systems described by Eisenhart~\cite{EisenhartClassical} as in the scalar case~\cite{Makarov1967}. (The list of systems coincides with the quantum mechanical results of \cite{Shapovalov1972}).

We deduce from these results that our system \eqref{super3int1radu} does not separate in any orthogonal coordinate system, as we cannot obtain the corresponding pair of integrals by combining the integrals $Y_i,H$ from \eqref{SI Y1}--\eqref{SI Y3} except the Cartesian one $Y_1$ and $X_2$ \eqref{SI X2}. As we have already shown, the system does not separate in this coordinate system either.
	
Moreover, the results from \cite{Benenti_2001} imply that the separation of Hamilton--Jacobi equation on any real Riemannian manifold is always orthogonal, so our superintegrable system~\eqref{super3int1radu} cannot separate in any choice of coordinates in the 3D Euclidean space. As far as we know, this is the first non--separable second order superintegrable system with a magnetic field.
 
This means that when a magnetic field is present, it is not possible to obtain all second order superintegrable systems by searching at the intersections of two separable systems nor two standard integrable ones, the ansatz used in \cite{Bertrand2020} inspired by Evans' paper \cite{Evans} from the natural case. Indeed, our system is not separable at all and has only one standard (Cartesian) pair of integrals.

Let us present a canonical transformation which reduces this system to a 2D one. We consider the generating function of the first type $F_1(x,\tilde{x})=b_1 (\tilde{y}-y) (\tilde{x}-x)$, i.e., we have the following transformation
\begin{equation}\label{2Dredtranf}
	\tilde{x} = -\frac{1}{b_1} p_y + x,\ \tilde{y} =-\frac{1}{b_1} p_x + y,\ \tilde{z} = z,\ p_{\tilde{x}} = p_x,\ p_{\tilde{y}} = p_y,\ p_{\tilde{z}} = p_z.
\end{equation}
In these coordinates, the Hamiltonian reads
\begin{equation}
	\mathscr{H}=\frac{1}{2}\left[ \left(p_{\tilde{x}} - \frac{b_3 a}{2}\cos\left(\frac{2\tilde{z}}{a}\right)\right)^2 +\left( b_1\tilde{x} +\frac{b_3 a}{2} \sin\left(\frac{2\tilde{z}}{a}\right)\right)^2 + p_{\tilde{z}}^2\right].
\end{equation}
It is therefore a 2D system with magnetic field
\begin{equation}
B_{2D}=b_3 \sin\left(\frac{2\tilde{z}}{a}\right) \mathrm{d}\tilde{x} \wedge \mathrm{d}\tilde{z}.
\end{equation}
It admits 2 trivial integrals, which coincide with the transformed first order integrals
\begin{equation}
	\tilde{Y}_1= b_1 \tilde{y},\quad \tilde{Y}_2 =p_{\tilde{y}}.
\end{equation}
The third integral $Y_3$ becomes a second order integral (simplified by subtracting $\tilde{Y}_1,\tilde{Y}_2$)
\begin{equation}
	\tilde{Y}_3=\tfrac{1}{2}(p_{\tilde{x}}^2 - b_1 a p_{\tilde{z}}+b_1^2 \tilde{x}^2),
\end{equation}
which yields integrability. The last integral $\tilde{X}_1$ is dependent on the Hamiltonian, so we omit it.

\begin{figure}
\centering
\includegraphics[scale=0.5]{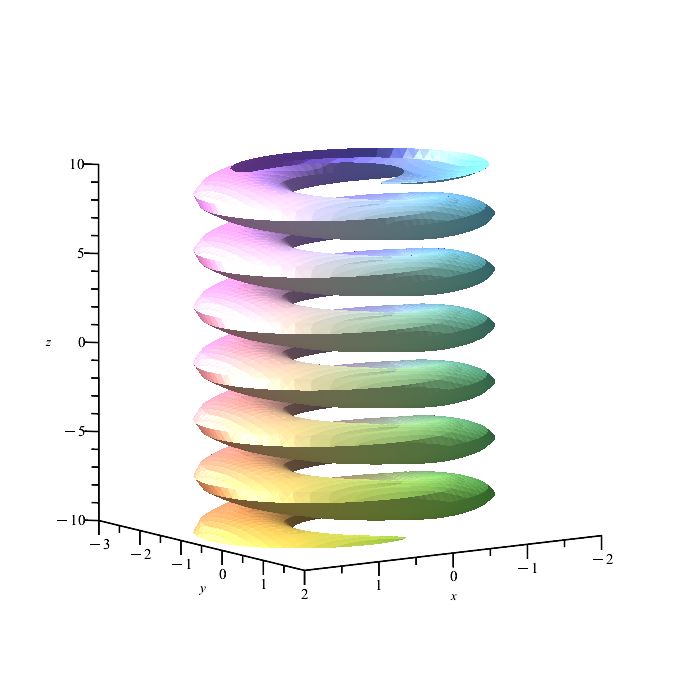}
\caption{Surface in the configuration space determined by the values of the integrals $E = \frac{11}{8} + \frac{\pi^2}{8} + \frac{\sqrt{3}}{2}, Y_1 = 1, Y_2 = 0, Y_3 = -\frac{1}{2} - \frac{\pi}{4}$ and the parameters $a = 1, b_1 = -\frac{\pi}{2}, b_3 = -\sqrt{3}$.}\label{Fig_surf}
\end{figure}

\begin{figure}
\centering
\includegraphics[scale=0.5]{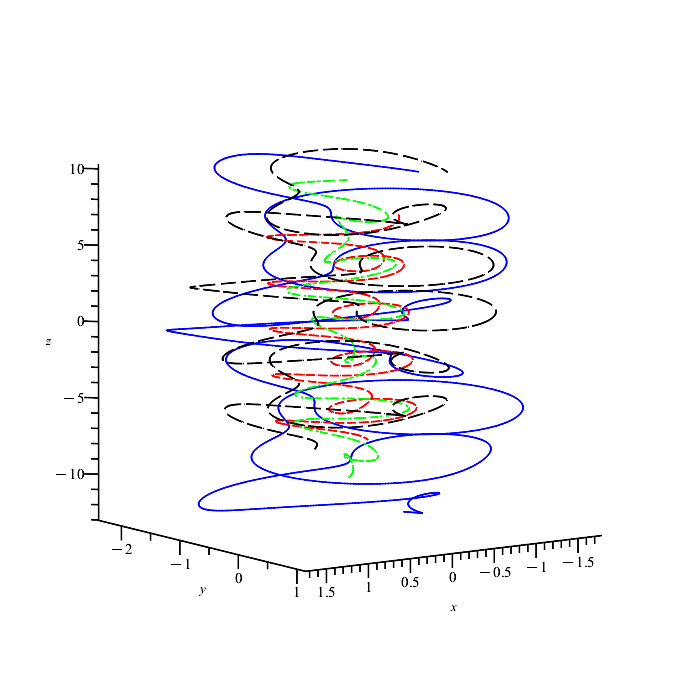}
\caption{Sample trajectories with the integrals $E = \frac{11}{8} + \frac{\pi^2}{8} + \frac{\sqrt{3}}{2}, Y_1 = 1, Y_2 = 0, Y_3 = -\frac{1}{2} - \frac{\pi}{4}$ and the parameters $a = 1, b_1 = -\frac{\pi}{2}, b_3 = -\sqrt{3}$.}\label{Fig_sampl_traj}
\end{figure}

As far as we know, the system~\eqref{super3int1radu} is minimally superintegrable. We did not succeed to solve the equations of motion for neither form of the system due to their complicated coupling. For given  values of the integrals $H=E,Y_1,Y_2,Y_3$ (cf. equations~\eqref{SI Y1}--\eqref{SI Y3}), its motion is restricted to a surface in the configuration space given by the equation
\begin{align}\label{surface}
2 E & =  \left(b_1 y+\frac{b_3 a}{2} \cos\left(\frac{2 z}{a}\right)  - Y_1\right)^2 + \left(b_1 x+\frac{b_3 a}{2} \sin\left(\frac{2 z}{a}\right)  + Y_2\right)^2 + \\ \nonumber 
& + \frac{1}{a^2}\left( b_1 \left(x^2 + y^2\right) + 2 x Y_2  - 2 y Y_1 - 2 Y_3\right)^2
\end{align}
which is periodic in the coordinate $z$ and quartic in the coordinates $x,y$. Due to the presence of the quartic terms the shape of this surface appears to be rather difficult to analyze analytically. For an example of it see Figure~\ref{Fig_surf}, numerically generated for the parameters $a = 1, b_1 = -\frac{\pi}{2}, b_3 = -\sqrt{3}$ and the values of the integrals $E = \frac{11}{8} + \frac{\pi^2}{8} + \frac{\sqrt{3}}{2}, Y_1 = 1, Y_2 = 0, Y_3 = -\frac{1}{2} - \frac{\pi}{4}$ which correspond to the initial conditions $x(0) = 1, y(0) = 0, z(0) = 0, p_x(0) = 1, p_y(0) = 0, p_z(0) = 1$. The motion inside this surface appears to be quite complicated, see Figure~\ref{Fig_sampl_traj} for several numerically plotted trajectories for the same value of the integrals. The trajectories are in general unbounded in the $z$ direction due to periodic form of the surface equation~\eqref{surface}; however, they are bounded in the $xy$ direction due to the dominant positive quartic terms in~\eqref{surface}.
 

Thus, there is no suggestion on how a hypothetical fifth independent integral which would make the system maximally superintegrable may look like. However, considering the results of~\cite{Marchesiello2015} for $b_1=0$ we expect that such an integral, if it exists, is non--polynomial in the momenta for $b_1\neq 0$, too. 

The fact that the classical trajectories are not bounded hints to expect that the quantum mechanical version of the system described by the fields~\eqref{super3int1radu} has continuous energy spectrum. Whether the four known integrals $H$, $Y_1$, $Y_2$ and $Y_3$ allow to give some more precise answers about the quantum spectrum is an open and rather nontrivial question going beyond the scope of the present paper.

We notice that in the absence of the electrostatic potential, $W(\vec x)=0$, the relativistic Hamiltonian expressed in the instant form (see, e.g.,~\cite{HeiIld}),
\begin{align*}
H_{\mathrm{rel}}=\sqrt{1+(p_j^A)^2}+W(\vec x),
\end{align*}
is a function of the nonrelativistic Hamiltonian~\eqref{HamMagn}. Thus the same nonabelian algebra of integrals of motion \eqref{SI Y1}--\eqref{SI Y3} is present also when motion of a relativistic electron in the helical undulator placed in a solenoid is considered, as long as the radiation emitted by the electron is neglected. This observation may be helpful in numerical modelling of undulators of the type proposed in~\cite{PhysRevAccelBeams.20.122401}.

\section{Generalized spherical case}\label{sec:GenSpher}

Let us now study the conditions for the existence of integrals of the form \eqref{int gen spher ext} with $b=0$, namely
\begin{equation}\label{GenSpherIntCart}
X_1=(\ell_3^{A})^2+\dots,\quad X_2=(\ell_1^{A})^2+ (\ell_2^{A})^2+ (\ell_3^{A})^2+ a \ell_3^{A} p_3^{A}+\dots
\end{equation}
As we search for generalized--type integrals, we assume $a\neq0$ in this section.

Notice that we could eliminate the $(\ell_3^{A})^2$ term from $X_2$ by subtracting $X_1$, however we prefer not to, as the equations in this way take simpler form.
In the spherical coordinates $(R,\theta,\phi)$ the integrals \eqref{GenSpherIntCart} read
\begin{align}\label{GenSpherInt}
	X_1={}& (p_{\phi}^{A})^2+ s_1^R p_R^{A}+ s_1^{\theta}p_{\theta}^{A}+ s_1^Z p_Z^{A} + m_1, \\
X_2={}& (p_{\theta}^{A})^2+\frac{(p_{\phi}^{A})^2}{\sin^2(\theta)}+a\left( p_R^{A}\, p_{\phi}^{A} \cos (\theta)-\frac{p_{\theta}^{A} p_{\phi}^{A} \sin (\theta)}{r} \right) + s_2^R p_R^{A}+ s_2^{\theta}p_{\theta}^{A}+ s_2^Z p_Z^{A} + m_2. \nonumber\\
\end{align}

Proceeding as in the generalized cylindrical case, we impose the integrability conditions \eqref{integrability} and start by looking at the resulting second order equations. We find
\begin{gather}
\partial_R s_1^R =0, \quad 
 s_1^R+R \partial_{\theta} s_1^{\theta}=0, \nonumber\\
 s_1^R+ R \left(\cot (\theta) s_1^{\theta}+ \partial_{\phi} s_1^{\phi}\right)=0,\quad 
\partial_{\theta} s_1^R+R^2 \partial_R s_1^{\theta}=0, \label{eq:X1-2ndOrder}\\
\partial_{\phi} s_1^R+( R^2\sin ^2(\theta))(\partial_{R} s_1^{\phi}-2B^{\theta})=0,\quad 
\partial_{\phi}s_1^{\theta}+ \sin^2(\theta)(\partial_{\theta} s_1^{\phi}+2 B^R)=0 \nonumber
\end{gather}
and
\begin{gather}
a \cos(\theta) B^{\theta}-\partial_R s_2^R=0,\quad a \sin (\theta) B^R- s_2^R-R \partial_{\theta}s_2^{\theta}=0, \nonumber\\
a(\sin(\theta)B^R+R \cos (\theta) B^{\theta})+ s_2^R+R( \cot (\theta) s_2^{\theta}+ \partial_{\phi} s_2^{\phi})=0, \nonumber\\
a( \cos (\theta) B^R+ R \sin(\theta)B^{\theta})+ \partial_{\theta}s_2^R +R^2 \partial_R s_2^{\theta}+2 R^2 B^{\phi}=0, \label{eq:X2-2ndOrder}\\
a R \sin^3 (\theta) B^{\phi}- \partial_{\phi} s_2^R-R^2 \left(\sin^2(\theta)\partial_R s_2^{\phi} -2 B^{\theta}\right)=0, \nonumber\\
a \cos (\theta)\sin^2(\theta) B^{\phi}-\partial_{\phi} s_2^{\theta}- \sin^{2}(\theta)\partial_{\theta} s_2^{\phi}=0 \nonumber
\end{gather}
for $X_1$ and $X_2$, respectively. The commutativity condition between the two integrals gives 
\begin{gather}
\partial_{\phi}s_1^R=0,\quad a \sin (\theta)\partial_{\phi}s_1^{\theta}-2 R \partial_{\theta} s_1^{\theta}=0,\quad a \left(R\cos (\theta)\partial_{\phi}s_1^{\theta}-\sin (\theta) \partial_{\phi} s_1^R\right)+2 R \partial_{\theta}s_1^R=0, \nonumber\\
2(\partial_{\phi} s_1^R-\sin^2(\theta)\partial_{\phi}s_2^R)\nonumber\\
+a\sin^2(\theta)\left(\sin(\theta)s_1^{\theta}+\cos\theta\partial_{\phi}s_1^{\phi}-\frac{\sin \theta}{R}\partial_{\theta}s_1^R+\cos\theta\partial_Rs_1^R\right)=0,\nonumber\\
\label{eq:X1X2-2ndOrder} a \sin^2(\theta) \left( R\cos(\theta)( \partial_R s_1^{\phi}-2B^{\theta})- \sin (\theta)(\partial_{\theta} s_1^{\phi}+ 2B^R)\right)\\
 +2 R(\cot (\theta) s_1^{\theta} + \partial_{\phi} s_1^{\phi}- \sin^2(\theta) \partial_{\phi} s_2^{\phi})=0, \nonumber\\
 a \,\sin^2(\theta) \left(
 R \cos(\theta) \left( R \partial_R s_1^{\theta}+s_1^{\theta} \right)-\sin (\theta) \left( s_1^R +R \partial_{\theta} s_1^{\theta} +R \partial_{\phi} s_1^{\phi} \right)
\right) \nonumber\\
+2 R^2 ( \partial_{\phi} s_1^{\theta}
+ \sin^2(\theta)(\partial_{\theta} s_1^{\phi}-\partial_{\phi} s_2^{\theta}+2 B^R))=0.\nonumber
\end{gather}
By solving \eqref{eq:X1-2ndOrder} we find
\begin{equation}\label{Bfirstsol}
 B^R=-\frac{1}{2} \left(\frac{1}{\sin^2(\theta)} \partial_{\phi}s_1^{\theta} +\partial_{\theta}s_1^{\phi}\right),\quad 
B^{\theta}= \frac12 \left(\partial_R s_1^{\phi}+\frac{\partial_{\phi}s_1^R}{ R^2\sin^2(\theta)}\right),
\end{equation}
where
\begin{align}
s_1^R={}&\cos (\theta) \partial_{\phi} S_{11}^R (\phi)-\sin (\theta) \partial_{\phi} S_{12}^R (\phi),\; 
s_1^{\theta}=\partial_{\phi} S_{1}^{\theta} (\phi)-\frac{\sin (\theta)\partial_{\phi} S_{11}^R (\phi)+\cos (\theta)\partial_{\phi} S_{12}^R (\phi)}{R},\nonumber
\\
s_1^{\phi}={}&S_{1}^{\phi}(R,\theta)+\frac{ S_{12}^R(\phi)}{R\sin(\theta)}-\cot (\theta) S_{1}^{\theta} (\phi).
\end{align}
By substituting this into \eqref{eq:X1X2-2ndOrder} and looking at the equations not containing $s_2^{R,\theta, \phi}$ we see that they are solved for
\begin{equation}
S_{11}^R(\phi)= s_{11}, \quad S_{12}^R(\phi)= s_{12},\quad S_1^{\theta}=s_{13},\quad s_{1j}\in\mathbb{R},\label{s11Sol}
\end{equation}
where we disregard solutions for which the magnetic field would not be periodic in $\phi$ nor have the same values at the boundaries $\phi=0$, $\phi=2\pi$. 
This implies
\begin{equation}
		s_1^R=s_1^{\theta}=0.
\end{equation}
After substituting \eqref{s11Sol} into \eqref{Bfirstsol} and replacing 
\begin{equation}
	S_1^{\phi}(R,\theta)=s_{13} \cot(\theta) -\frac{F(R,\theta)}{R}-\frac{s_{12}}{R\sin(\theta)},
\end{equation}
we arrive at
\begin{equation}\label{BF}
B^R= \frac{\partial_{\theta} F(R,\theta)}{2 R},\quad B^{\theta}= \frac{F(R,\theta)-R \partial_R F(R,\theta)}{2 R^2}
\end{equation}
and
\begin{equation}
	s_1^{\phi}=-\frac{F(R,\theta)}{R}.
\end{equation}
The remaining equations in \eqref{eq:X1X2-2ndOrder} are then solved by
\begin{equation}\label{s2firstsol}
s_2^R= S_2^R(R,\theta),\quad s_2^{\theta}= S_2^{\theta}(R,\theta),\quad s_2^{\phi}= S_2^{\phi}(R,\theta).
\end{equation}
Furthermore, by solving the last equation in \eqref{eq:X2-2ndOrder} we obtain
\begin{equation}\label{BphiSol}
B^{\phi}= \frac{\partial_{\theta}S_2^{\phi}(R,\theta)}{a \cos(\theta)}.
\end{equation}
At this point the second order equations for $X_2$ are still not fully solved. However, by looking at the zero-order equation for $X_1$ we realize that it reduces to
\begin{equation}
	F(R,\theta) \partial_{\phi} W=0.
\end{equation}
If $F(R,\theta)=0$, the only non--zero component of the magnetic field remains $B^{\phi}$. However, by substituting \eqref{s2firstsol}, \eqref{BF}, \eqref{BphiSol} and  $F=0$ into \eqref{eq:X2-2ndOrder} we find that necessarily also $B^{\phi}=0$.
Since we are interested in solutions for nonvanishing magnetic field, we conclude that $W= W(R, \theta)$. 

Thus, the magnetic field and the electrostatic potential do not depend on $\phi$ anymore and indeed, by a straightforward calculation, one can see that the remaining equations for $X_1$ are solved by $X_1=p_{\phi}^2$ if the vector potential corresponding to the magnetic field~\eqref{BF} and~\eqref{BphiSol} is chosen independent of $\phi$, i.e., in the form
\begin{equation}\label{gaugeSpher}
A_{R}=\tilde{A}_{R}(R,\theta),\quad A_{\theta}=\tilde{A}_{\theta}(R,\theta),\quad A_{\phi}=\frac{F(R,\theta)}{2R}.
\end{equation}
This implies that the commutativity condition of the two integrals is solved by
\begin{equation}
m_2= M_{2}(R,\theta).
\end{equation}
Thus, not only the magnetic field and the potential, but also both integrals (in proper gauge choice) do not depend on $\phi$. By setting $p_{\phi}$ equal to a constant $p_0$, we can reduce to the problem of the existence of an integral of the type
\begin{equation}\label{polarX2}
X_2=(p_{\theta}^{A})^2+\ldots
\end{equation}
for the system in two degrees of freedom determined by the magnetic field
 \begin{equation}\label{2DmagfieldSpher}
B_{2D}=\frac{\partial_{\theta}S_2^{\phi}}{a \cos(\theta)} {\mathrm d} R \wedge {\mathrm d} \theta
\end{equation}
and the Hamiltonian
\begin{equation}\label{2DHspher}
H_{2D}=\frac12\left((p_R^A)^2+ \frac{(p_{\theta}^A)^2}{R^2}\right)+ \frac{1}{2R^2\sin^2\theta}\left(p_0+\frac{ F(R,\theta)}{2R}\right)^2+ W(R,\theta)
\end{equation}
in the coordinates $(R,\theta)$. Notice that in order to obtain the above reduction we had to choose a gauge, namely~\eqref{gaugeSpher}.
Therefore, the potential in \eqref{2DHspher} depends on the gauge of the original 3D system.

Moreover, for $\partial_{\theta}S_2^{\phi}=0$ such reduced system does not contain magnetic field anymore. In this case the remaining equations for $X_2$ can be easily solved. We find only one system that can satisfy them, with the magnetic field
\begin{equation}\label{sphintsys}
B^R=\frac{\kappa_1}{2} R^2\sin(2\theta),\quad B^{\theta}=-\kappa_1 R\sin^2(\theta),\quad B^{\phi}=0,\quad \kappa_1 \in\mathbb R,
\end{equation}
which in Cartesian coordinates is constant,
\begin{equation}
\vec B(\vec x)=\left( 0,0,\kappa_1\right),
\end{equation}
and
\begin{equation}
W(R,\theta)=-\frac{\kappa_1^2}{8}R^2\sin^2(\theta)-\frac{\lambda_1}{R^2\sin^2(\theta)} =-\frac{\kappa_1^2}{8}(x^2+y^2)-\frac{\lambda_1}{x^2+y^2}.
\end{equation}
This system is not new, it is already known \cite{Marchesiello2018Sph} to possess standard spherical--type integrals (i.e., $a=0$ in \eqref{GenSpherIntCart}), together with one cylindrical--type integral, namely
\begin{equation}
	X_3= \cos (u) p_R^A-\frac{\sin (u) p_{\theta}^A}{R}=p_Z^A,
\end{equation}
that in Cartesian coordinates and proper gauge choice simplifies to $p_3$. Thus, the term proportional to $a$ in \eqref{GenSpherIntCart} is a constant of motion.

The only remaining possibility is that $\partial_{\theta} S_2^{\phi}\neq0$ so that the reduced system still has magnetic field. In this case, we observe that the Hamiltonian \eqref{2DHspher} looks as if written in polar coordinates. However, due to the reduction from a three dimensional system expressed in the spherical coordinates, we have $\theta\in[0,\pi]$ and not $\theta\in[0,2\pi)$ as for the angular variable in polar coordinates,  i.e., $(R,\theta)$ parameterize only the half--plane instead of the full $xy$--plane. Nevertheless, in our considerations we never use any assumption on the range of the variable $\theta$, and the remaining equations can be solved formally as if working in polar coordinates. Thus, we reduce to the search of a polar--type integral \eqref{polarX2} in presence of a magnetic field in polar coordinates. Also in \cite{Berube}, where the problem was already studied, the assumption on the range of the angle variable is not explicitly needed in any of the steps.

Therefore, following \cite{Berube}, we see that we can simplify our problem as there can be only two possibilities for the magnetic field, namely
\begin{enumerate}
\item $ B^{\phi}$ depends only on $R$, therefore
\begin{equation}\label{S2simpl1}
S_2^{\phi}=S_{21}^{\phi}(R)+a \sin (\theta) \partial_R S_{22}^{\phi}(R),
\end{equation}
or
\item
\begin{equation}\label{BSimpl2}
 B^{\phi}= -\frac{A+\partial_{\theta}^2 S(\theta)+ S(\theta)}{2 R^3}, \quad A\in\mathbb R
 \end{equation}
where the function $S$ is not arbitrary but has to satisfy
\begin{equation} \label{Scond}
(S+ A)^2\partial_{\theta}\left( \frac{\partial^3_{\theta}S }{\partial_{\theta} S}\right)+ 3 (S+ A)(8\partial_{\theta} S + 3\partial^3_{\theta} S) + 15 \partial_{\theta} S\partial^2_{\theta} S=0.
\end{equation}

\end{enumerate}

In both cases, we find it convenient to simplify the second order equations for $X_2$ by the substitution
\begin{equation}
S_2^R(R,\theta)=G^R (R,\theta)-\frac{a \cos (\theta) F(R,\theta)}{2 R},\quad S_2^{\theta}(R,\theta)= G^{\theta}(R,\theta)+\frac{a \sin (\theta) F(R,\theta)}{2 R^2},
\end{equation}
so that they read
\begin{gather}
\partial_R G^R=0,\quad G^R+ R\partial_{\theta}G^{\theta}=0,\nonumber\\
a\sin(\theta)( \partial_{\theta} F- R \cot (\theta) \partial_R F + \cot (\theta) F) +2 R G^R+2 R^2 \cot (\theta) G^{\theta}=0, \nonumber\\
\frac{2 \partial_{\theta} S_2^{\phi}}{a \cos(\theta)}+\frac{\partial_{\theta} G^R}{R^2}+\partial_R G^{\theta}=0,\label{2ndord2equiv}\\
 F-R \left(\partial_RF-\sin^2(\theta) \tan(\theta) \partial_{\theta}S_2^{\phi} +R\sin^2(\theta)\partial_RS_2^{\phi}\right)=0.\nonumber
\end{gather}

By imposing \eqref{S2simpl1} or \eqref{BSimpl2}, a straightforward computation leads to the conclusion that the above equations can be solved only if $B^{\phi}=0$. Namely, in the first case we arrive at 
 $S_{22}^{\phi}=b_{1}$ and $S_{21}^{\phi}=\frac{1}{2} b_{2} R^2+b_{3}$, $b_j\in\mathbb R$. This implies $B^{\phi}=0$.
In the second case, it is even enough to impose the numerator in $B^{\phi}$ to be function only of $\theta$, by setting 
\begin{equation}
	S_2^{\phi}=S_{21}(R)+\frac{ a S_{22}(\theta)}{R^3},
\end{equation}
without the further constraint \eqref{Scond}, to conclude that $S_{21}=c_1 R^2-a c_2 R^{-3}+c_3$ and $S_{22}=c_2$, with $c_j\in\mathbb R$, and thus $B^{\phi}=0$. 

Therefore, we conclude that no new system in a nonvanishing magnetic field can be found with two commuting integrals in the form \eqref{GenSpherIntCart}.\medskip

An alternative way to the same conclusion, avoiding the reduction to the 2D problem, is also feasible, albeit computationally more complicated. From the second order equations for $X_2$ expressed in the form~\eqref{2ndord2equiv} one finds 
\begin{align}
\nonumber G^R &=\frac{{\mathrm d}g^R(\theta)}{{\mathrm d}\theta}, \qquad G^\theta=-\frac{g^R(\theta)}{R}+g^\theta(R), \\
 S_2^\phi &=-\frac{a \sin(\theta)}{2 R^2} g^R(\theta)-\frac{a \cos(\theta)}{2 R^2} \frac{{\mathrm d} g^R(\theta)}{{\mathrm d} \theta}-\frac{a \sin(\theta)}{2} \frac{{\mathrm d} g^\theta(R)}{{\mathrm d} R}+h(R).
\end{align}
 The remaining two equations in~\eqref{2ndord2equiv} can be solved with respect to the first order derivatives of $F$. Their Clairaut's compatibility condition $\partial_\theta(\partial_R F(R,\theta))=\partial_R(\partial_\theta F(R,\theta)) $ involves only undetermined functions of single variable $g^R(\theta)$, $g^\theta(R)$ and $h(R)$, and thus implies a system of ODEs for them. Solving it and next determining $F(R,\theta)$ one can find the most general solution of the second order equations~\eqref{2ndord2equiv}, depending on 8 integration constants. These are in turn constrained by the remaining lower order equations, leading to the unique system~\eqref{sphintsys} under the assumption of nonvanishing magnetic field. 

\section{Conclusions}\label{sec:con}
We have demonstrated that only one system can be found in a nonvanishing magnetic field with two commuting integrals in the form \eqref{GenSpherIntCart}, however this system admits a pair of standard spherical--type integrals. When considering integrable systems with generalized cylindrical--type integrals \eqref{int ext cyl}, the results are richer and we found three new systems with such integrals, see \eqref{system c<>0}, \eqref{xi system} and \eqref{solutionc=0b3<>0}.

Integrability with generalized Cartesian--type integrals, i.e., integrals of the form
\begin{equation}
	X_1= p_1^2 + a p_2^2,\quad X_2=p_2^2+\sum_{i,j} b_{ij} p_i p_j+\ldots,\quad a,b_{ij}\in\mathbb R,\;i\neq j,
\end{equation}
is also possible \cite{Marchesiello2017}, as well as with integrals of the form \cite{Marchesiello2022}
\begin{equation}\label{intgen}
 X_1=\ell_3p_3+ a p_1^2+\dots ,\quad X_2=p_3,\quad a\neq0
\end{equation}
that do not correspond to any of the known classes of admissible commuting integrals in absence of magnetic field~\cite{Makarov1967}. Thus, when vector potentials are present, integrals can have more general forms with respect to the case with only scalar potentials. 

However, for all the examples known so far, the systems with generalized--type integrals separate in cylindrical or in Cartesian coordinates in the limit of vanishing magnetic field, even when generalized spherical--type integrals or even more general integral of the form~\eqref{intgen} are considered. Thus, Cartesian-- and cylindrical--type integrals seem to play some distinguished role in allowing possible deformations that preserve integrability, albeit not separability, in a magnetic field.

Cartesian and cylindrical coordinates are subgroup type coordinates, related to the two maximal abelian Lie subalgebras of the three dimensional Euclidean algebra $\mathfrak{e}(3)$, namely $\mathcal A_1=\mathrm{span}\{p_3,\ell_3\}$ and $\mathcal A_2=\mathrm{span}\{p_1,p_2,p_3\}$, respectively. Spherical coordinates are subgroup type coordinates as well, however to have commuting elements with the kinetic term of the Hamiltonian, the corresponding integrals have leading order terms in the enveloping algebra of $\mathfrak{e}(3)$, namely $\mathcal A_3=\{\ell_3, L^2\}$, where $L^2=\sum_j\ell_j^2$, i.e., one term is quadratic.
All other orthogonal coordinate systems in the 3D Euclidean space are not subgroup type and are related to integrals with second order terms that do not belong to these subalgebras. 

In the absence of magnetic field there is some disparity as well between subgroup and non--subgroup type coordinates in the possible allowed form of independent integrals. However, it first appears when considering at least one third order integral \cite{Marchesiello2015a}. In the presence of magnetic fields such disparity seems to appear already with second order integrals. The presence of linear terms in the Hamiltonian may be the reason why separation in the limit of vanishing vector potentials is possible only in the subgroup type coordinates related to the subalgebras $\mathcal A_1$ and $\mathcal A_2$. Understanding the mechanism behind this behavior may be a key towards the complete classification of quadratic integrable systems in magnetic fields and  we shall pursue its investigation in future work.

We have found only one new superintegrable system among the generalized cylindrical--type integrable ones which lies at the intersection of Cartesian--type and generalized cylindrical--type of integrals. We deem this rarity to be caused by the rather restricted form of such integrable systems. In contrast to the systems with standard form of integrals, whose magnetic field and electrostatic potential contain arbitrary functions of single variable, in the cases with generalized--type integrals only real--valued constants remain unconstrained. This does not leave much freedom for the existence of additional integrals.
		 
The superintegrable system found here is the first known non--separable one and contains generalized integrals nonreducible to standard ones. As we have already mentioned, this means that when a magnetic field is present, it is not possible to obtain all second order superintegrable systems by looking at the intersection of two separable systems nor two standard integrable ones which was the ansatz used in \cite{Bertrand2020} inspired by Evans' paper \cite{Evans} from the scalar case. To find all quadratically superintegrable systems with magnetic field, we have to search beyond the separable (and also standard integrable) cases.

There is a conjecture, born out by all known examples, that all maximally superintegrable systems are exactly solvable \cite{TTW}. For quantum systems this means that the energy levels can be calculated algebraically and the wave functions are polynomials multiplied by a gauge factor, in appropriately chosen variables. If the system is separable, such coordinates are naturally chosen as the ones related to separation. To our knowledge, even with a magnetic field, no maximally superintegrable system has been found to refute such a conjecture. All the known examples of maximally superintegrable systems are either separable \cite{Kubu_2021,Marchesiello2019}, or can be reduced to a separable one \cite{Marchesiello2018} by canonical transformation similar to \eqref{2Dredtranf}. The new superintegrable system we presented in this paper appears to be only minimally superintegrable, therefore it can not give any further insight in this direction.

It remains a question on how to approach (super)integrable but not separable systems (in the classical sense, i.e. separation of the Hamilton--Jacobi equation in the configuration space). E.g. our system \eqref{super3int1radu} seems not to possess periodic trajectories for generic values of the initial data, thus complicating the search for the canonical transformation to action--angle coordinates (in which the system would separate in the phase space). Here the approach employing Haantjes algebras \cite{NoTemTo} may prove helpful. In some cases, the integrals may allow quasi-separation techniques \cite{Charest2007} or block--separation of variables \cite{ChaRas}. We postpone the study of this problem for future work.

For physical applications, the results of the Section~\ref{nonsep Cart} are of particular interest. Our observation that the electron in helical undulator placed in a solenoid possesses a nontrivial algebra of integrals of motion may help in the study of its properties, e.g., in allowing more efficient numerical simulations, but this goes far beyond the scope and goals of the present paper.

\section*{Acknowledgment}
OK was supported by the Grant Agency of the Czech Technical University in Prague, grant No. SGS22/178/OHK4/3T/14. Research of AM was partially supported by GNFM--INdAM. L\v{S} was supported by the project of the Ministry of Education, Youth and Sports of the Czech Republic \verb|CZ.02.1.01/0.0/0.0/16_019/0000778| Centre of Advanced Applied Sciences, co--financed by the European Union.

Computations in this paper were performed using Maple\textsuperscript{TM} 2021.2 by Maplesoft, a division of Waterloo Maple Inc., Waterloo, Ontario and 
Mathematica, Version 13.0.0 by Wolfram Research, Inc., Champaign, IL.

\pdfbookmark[1]{References}{ref}

\begin{thebibliography}{10}
\expandafter\ifx\csname url\endcsname\relax
  \def\url#1{\texttt{#1}}\fi
\expandafter\ifx\csname urlprefix\endcsname\relax\def\urlprefix{URL }\fi
\expandafter\ifx\csname href\endcsname\relax
  \def\href#1#2{#2} \def\path#1{#1}\fi

\bibitem{Makarov1967}
A.~Makarov, J.~Smorodinsky, K.~Valiev, P.~Winternitz, A systematic search for
  nonrelativistic systems with dynamical symmetries, Nuovo Cimento A Series 10
  (1967) 1061--1084.
\newblock \href {http://dx.doi.org/10.1007/BF02755212}
  {\path{doi:10.1007/BF02755212}}.

\bibitem{Evans}
N.~W. Evans, Superintegrability in classical mechanics, Physical Review A
  41~(10) (1990) 5666--5676.
\newblock \href {http://dx.doi.org/10.1103/PhysRevA.41.5666}
  {\path{doi:10.1103/PhysRevA.41.5666}}.

\bibitem{Marchesiello2022}
A.~Marchesiello, L.~{\v{S}}nobl, Pairs of commuting quadratic elements in the
  universal enveloping algebra of {E}uclidean algebra and integrals of motion,
  Journal of Physics A: Mathematical and Theoretical 55~(14) (2022) 145203
  \newblock\href{http://dx.doi.org/10.1088/1751-8121/ac515e}
  {\path{doi:10.1088/1751-8121/ac515e}}.

\bibitem{Marchesiello2017}
A.~Marchesiello, L.~\v{S}nobl, Superintegrable 3{D} systems in a magnetic field
  corresponding to {C}artesian separation of variables, Journal of Physics A: Mathematical and Theoretical
  50~(24) (2017) 245202.
\newblock \href {http://dx.doi.org/10.1088/1751-8121/aa6f68}
  {\path{doi:10.1088/1751-8121/aa6f68}}.

\bibitem{Shapovalov1972}
V.~N. Shapovalov, V.~G. Bagrov, A.~G. Meshkov, Separation of variables in the
  stationary {S}chroedinger equation, Soviet Physics Journal [translation of
  Izvestiya Vysshikh Uchebnykh Zavedenii, Fizika] 15~(8) (1972) 1115--1119.
\newblock \href {http://dx.doi.org/10.1007/bf00910289}
  {\path{doi:10.1007/bf00910289}}.

\bibitem{Benenti_2001}
S.~Benenti, C.~Chanu, G.~Rastelli, Variable separation for natural
  {H}amiltonians with scalar and vector potentials on {R}iemannian manifolds,
  Journal of Mathematical Physics 42~(5) (2001) 2065--2091.
\newblock \href {http://dx.doi.org/10.1063/1.1340868}
  {\path{doi:10.1063/1.1340868}}.

\bibitem{KMW76}
E.~Kalnins, W.~Miller, Jr., P.~Winternitz, The group {${\rm O}(4)$}, separation
  of variables and the hydrogen atom, SIAM Journal on Applied Mathematics
  30~(4) (1976) 630--664.
\newblock \href {http://dx.doi.org/10.1137/0130058}
  {\path{doi:10.1137/0130058}}.

\bibitem{MPW81}
W.~Miller, Jr., J.~Patera, P.~Winternitz, Subgroups of {L}ie groups and
  separation of variables, Journal of Mathematical Physics 22~(2) (1981)
  251--260.
\newblock \href {http://dx.doi.org/10.1063/1.524896}
  {\path{doi:10.1063/1.524896}}.

\bibitem{Marchesiello2015a}
A.~Marchesiello, S.~Post, L.~{\v{S}}nobl, Third-order superintegrable systems
  with potentials satisfying only nonlinear equations, Journal of Mathematical
  Physics 56~(10) (2015) 102104.
\newblock \href {http://dx.doi.org/10.1063/1.4933218}
  {\path{doi:10.1063/1.4933218}}.

\bibitem{Marchesiello2015}
A.~Marchesiello, L.~\v{S}nobl, P.~Winternitz, Three-dimensional superintegrable
  systems in a static electromagnetic field, Journal of Physics A: Mathematical and Theoretical 48~(39)
  (2015) 395206.
\newblock \href {http://dx.doi.org/10.1088/1751-8113/48/39/395206}
  {\path{doi:10.1088/1751-8113/48/39/395206}}.

\bibitem{Miller2013}
W.~Miller, Jr., S.~Post, P.~Winternitz, Classical and quantum
  superintegrability with applications, Journal of Physics A: Mathematical and Theoretical 46~(42) (2013)
  423001.
\newblock \href {http://dx.doi.org/10.1088/1751-8113/46/42/423001}
  {\path{doi:10.1088/1751-8113/46/42/423001}}.

\bibitem{Fournier2019}
F.~Fournier, L.~\v{S}nobl, P.~Winternitz, Cylindrical type integrable classical
  systems in a magnetic field, Journal of Physics A: Mathematical and
  Theoretical 53~(8) (2020) 085203.
\newblock \href {http://dx.doi.org/10.1088/1751-8121/ab64a6}
  {\path{doi:10.1088/1751-8121/ab64a6}}.

\bibitem{Kubu_2021}
O.~Kub{\r{u}}, A.~Marchesiello, L.~{\v{S}}nobl, Superintegrability of separable
  systems with magnetic field: the cylindrical case, Journal of Physics A:
  Mathematical and Theoretical 54~(42) (2021) 425204.
\newblock \href {http://dx.doi.org/10.1088/1751-8121/ac2476}
  {\path{doi:10.1088/1751-8121/ac2476}}.

\bibitem{Marchesiello2019}
A.~Marchesiello, L.~\v{S}nobl, Classical superintegrable systems in a magnetic
  field that separate in {C}artesian coordinates, SIGMA. Symmetry,
  Integrability and Geometry. Methods and Applications 16 (2020) 015.
\newblock \href {http://dx.doi.org/10.3842/SIGMA.2020.015}
  {\path{doi:10.3842/SIGMA.2020.015}}.

\bibitem{Kubu2020}
O.~Kub\r{u}, \href{http://hdl.handle.net/10467/90871}{Integrable and
  superintegrable systems of cylindrical type in magnetic fields}, Master's
  thesis, Czech Technical University in Prague, Prague (2020).
\newblock \href {http://arxiv.org/abs/2210.02393}
  {\path{doi:10.48550/arXiv.2210.02393}}.


\bibitem{Landau}
L.~D. Landau, E.~M. Lifshitz, Quantum mechanics: non-relativistic theory.
  {C}ourse of {T}heoretical {P}hysics, {V}ol. 3, Addison-Wesley Series in
  Advanced Physics, Pergamon Press Ltd., London-Paris, 1958, translated from
  the Russian by J. B. Sykes and J. S. Bell.

\bibitem{HeiIld}
T.~Heinzl, A.~Ilderton, Superintegrable relativistic systems in
  spacetime-dependent background fields, Journal of Physics A: Mathematical and Theoretical 50~(34) (2017)
  345204, 14.
\newblock \href {http://dx.doi.org/10.1088/1751-8121/aa7fa3}
  {\path{doi:10.1088/1751-8121/aa7fa3}}.

\bibitem{PhysRevAccelBeams.20.122401}
N.~Balal, I.~V. Bandurkin, V.~L. Bratman, A.~E. Fedotov, Helical undulator
  based on partial redistribution of uniform magnetic field, Physical Review
  Accelerators and Beams 20 (2017) 122401.
\newblock \href {http://dx.doi.org/10.1103/PhysRevAccelBeams.20.122401}
  {\path{doi:10.1103/PhysRevAccelBeams.20.122401}}.

\bibitem{Snobl_book}
L.~\v{S}nobl, P.~Winternitz, {C}lassification and {I}dentification of {L}ie
  {A}lgebras, Vol.~33 of CRM Monograph series, American Mathematical Society,
  Providence, Rhode Island, 2014.

\bibitem{Zhalij_2015}
A.~Zhalij, Quantum integrable systems in three-dimensional magnetic fields: the
  {C}artesian case, Journal of Physics: Conference Series 621 (2015) 012019.
\newblock \href {http://dx.doi.org/10.1088/1742-6596/621/1/012019}
  {\path{doi:10.1088/1742-6596/621/1/012019}}.

\bibitem{LeviCivita1904}
T.~Levi-Civita, Sulla integrazione della equazione di {H}amilton-{J}acobi per
  separazione di variabili, Mathematische Annalen 59~(3) (1904) 383--397.
\newblock \href {http://dx.doi.org/10.1007/bf01445149}
  {\path{doi:10.1007/bf01445149}}.

\bibitem{EisenhartClassical}
L.~P. Eisenhart, Separable systems in {E}uclidean 3-space, Physical Review (2)
  45 (1934) 427--428.
\newblock \href {http://dx.doi.org/https://doi.org/10.1103/PhysRev.45.427.2}
  {\path{doi:https://doi.org/10.1103/PhysRev.45.427.2}}.

\bibitem{Bertrand2020}
S.~Bertrand, O.~Kub{\r{u}}, L.~{\v{S}}nobl, On superintegrability of 3{D}
  axially-symmetric non-subgroup-type systems with magnetic fields, Journal of
  Physics A: Mathematical and Theoretical 54~(1) (2020) 015201.
\newblock \href {http://dx.doi.org/10.1088/1751-8121/abc4b8}
  {\path{doi:10.1088/1751-8121/abc4b8}}.

\bibitem{Marchesiello2018Sph}
A.~Marchesiello, L.~\v{S}nobl, P.~Winternitz, Spherical type integrable
  classical systems in a magnetic field, Journal of Physics A: Mathematical and Theoretical 51~(13) (2018)
  135205.
\newblock \href {http://dx.doi.org/10.1088/1751-8121/aaae9b}
  {\path{doi:10.1088/1751-8121/aaae9b}}.

\bibitem{Berube}
J.~B\'{e}rub\'{e}, P.~Winternitz, Integrable and superintegrable quantum
  systems in a magnetic field, Journal of Mathematical Physics 45~(5) (2004)
  1959--1973.
\newblock \href {http://dx.doi.org/10.1063/1.1695447}
  {\path{doi:10.1063/1.1695447}}.

\bibitem{TTW}
P. Tempesta,  V. Turbiner, P. Winternitz, Exact solvability of superintegrable systems Journal of Mathematical Physics 42 (2001) 4248.
\newblock \href {https://doi.org/10.1063/1.1386927}
  {\path{doi.org/10.1063/1.1386927}}.

\bibitem{Marchesiello2018}
A.~Marchesiello, L.~\v{S}nobl, An infinite family of maximally superintegrable systems in a magnetic field with higher order integrals, SIGMA. Symmetry, Integrability and Geometry: Methods and Applications 14 (2018) 092.
\newblock \href {http://doi.org/10.3842/SIGMA.2018.092}
  {\path{doi:10.3842/SIGMA.2018.092}}.

\bibitem{NoTemTo}
D. Reyes Nozaleda, P. Tempesta, G. Tondo, Classical multiseparable Hamiltonian systems, superintegrability and Haantjes geometry, Communications in Nonlinear Science and Numerical Simulation 104 (2022) 106021.
\newblock \href {http://doi.org/10.1016/j.cnsns.2021.106021}
  {\path{doi:10.1016/j.cnsns.2021.106021}}.

\bibitem{Charest2007}
F. Charest, C. Hudon, P. Winternitz, Quasiseparation of variables in the Schr\"{o}dinger equation with a magnetic field, Journal of Mathematical Physics 48 (2007) 012105.
\newblock \href {http://doi.org/10.1063/1.2399087}
  {\path{doi:10.1063/1.2399087}}.

\bibitem{ChaRas}
C. M. Chanu, G. Rastelli, Block-separation of variables: a form of partial separation for natural Hamiltonians, 
SIGMA. Symmetry, Integrability and Geometry: Methods and Applications 15 (2019) 013.
\newblock \href {http://www.emis.de/journals/SIGMA/2019/013/}
  {\path{doi:10.3842/SIGMA.2019.013}}.



\end{thebibliography}

\end{document}